\documentclass[%
aip,
amsmath,amssymb,
preprint,%
]{revtex4-1}

\usepackage{graphicx}% Include figure files
\usepackage{dcolumn}% Align table columns on decimal point
\usepackage{bm}% bold math
\usepackage[utf8]{inputenc}
\usepackage[T1]{fontenc}
\usepackage{mathptmx}
\usepackage{etoolbox}

\makeatletter
\def\@email#1#2{%
 \endgroup
 \patchcmd{\titleblock@produce}
  {\frontmatter@RRAPformat}
  {\frontmatter@RRAPformat{\produce@RRAP{*#1\href{mailto:#2}{#2}}}\frontmatter@RRAPformat}
  {}{}
}%
\makeatother
\begin{document}

%\preprint{AIP/123-QED}

\title[CO\textsubscript{2} Dissociative Sticking on Cu(110)]{CO\textsubscript{2} Dissociative Sticking on Cu(110)}
% Force line breaks with \\
\author{Federico J. Gonzalez}%
\affiliation{ Grupo de Fisicoqu\'imica en Interfaces y Nanoestructuras, Instituto de F\'isica Rosario (IFIR), CONICET-UNR, Bv. 27 de Febrero 210 bis, S2000EKF Rosario, Argentina}
\author{Carmen A. Tachino}%
\affiliation{ Grupo de Fisicoqu\'imica en Interfaces y Nanoestructuras, Instituto de F\'isica Rosario (IFIR), CONICET-UNR, Bv. 27 de Febrero 210 bis, S2000EKF Rosario, Argentina}%
\affiliation{Facultad de Ciencias Exactas, Ingenier\'ia y Agrimensura, Universidad Nacional de Rosario, Av. Pellegrini 250,
S2000 Rosario, Argentina}%
\author{H. Fabio Busnengo}%
 %\email[Corresponding author: ]{busnengo@ifir-conicet.gov.ar}
\affiliation{ Grupo de Fisicoqu\'imica en Interfaces y Nanoestructuras, Instituto de F\'isica Rosario (IFIR), CONICET-UNR, Bv. 27 de Febrero 210 bis, S2000EKF Rosario, Argentina}%
\affiliation{Facultad de Ciencias Exactas, Ingenier\'ia y Agrimensura, Universidad Nacional de Rosario, Av. Pellegrini 250,
S2000 Rosario, Argentina}%
\email[Corresponding author: ]{busnengo@ifir-conicet.gov.ar}

\date{\today}% It is always \today, today,

\begin{abstract}
In this work we investigate the dissociation of CO$_2$ on Cu(110)
by performing density functional theory calculations using the vdW-DF2 exchange-correlation functional. 
The total energies obtained were employed to parameterize an 
artificial neural network potential energy surface by using 
an active learning iterative approach.
The obtained potential was then used in quasi-classical trajectory calculations of molecular and dissociative adsorption probabilities
as a function of the initial impact energy of the molecules and surface temperature. 
We compare our results with available supersonic molecular beam experimental data for normal incidence.
Concerning the general dependence of the molecular and dissociative adsorption probabilities on the initial translational energy of the molecules, our theoretical results agree with experiments.
Also in agreement with experiments we have found that dissociative adsorption is not affected by surface temperature between 50 K and 400 K, for impact energies for which the dissociation probability is larger than $\sim 10^{-3}$.
We have investigated the influence of impact energy and surface temperature on the final state of the dissociation products by extending the time integration of the reactive trajectories up to
10 ps. We have found that above $\sim 2.5$~eV and close to or above room temperature, CO$_2$ dissociation induces strong surface distortions including final structures involving Cu adatoms. The creation of Cu vacancy-adatom pairs is stimulated by the presence of both CO$_{\mathrm{ads}}$ and O$_{\mathrm{ads}}$ which interact strongly with the Cu adatoms and even give rise to unexpected (O-Cu-CO)$_{\mathrm{ads}}$ linear moieties anchored to the surface by the dissociated O atom and involving a Cu adatom almost detached from the surface.
These surface distortions produced by dissociation products of high
energy CO$_2$ molecules at and above room temperature might explain
recent experiments that have found a saturation oxygen coverage 
for high energy molecules, larger than for slow molecules.
\end{abstract}

\maketitle

\section{Introduction}

The interaction of CO$_2$ with copper surfaces has attracted considerable attention due to the application of Cu-based catalysts in the conversion of CO$_2$ to value-added products, such as methanol, ethanol, and other C$_1$ and C$_{2+}$ compounds \cite{Gattrell2006,Kondratenko2013,Niu2022}. 
A fundamental challenge in this process is the activation of CO$_2$ due to its high stability\cite{Zhu2019,Kass2024}. 
Though the most efficient Cu-based catalysts for CO$_2$ conversion
consist of Cu particles deposited on oxide supports\cite{Kattel2017,Guzman2022}, understanding in detail the way CO$_2$ interacts and adsorbs on simpler systems such as flat Cu surfaces is also of great interest.
Whether CO$_2$ dissociates or not upon adsorption on Cu(110) 
(the most active low-Miller-index face of copper\cite{Wang2004})
has been a subject of debate.
Early ultra-high vacuum (UHV) experimental studies found that low energy CO$_2$ molecules do not chemisorb on Cu(110) and the estimated energy barrier ($\mathrm{E_{b}}$) for dissociation was 0.69 eV\cite{Nakamura1989,Fu1992}.
However, experiments conducted by Funk \textit{et al.} with supersonic molecular beams (SMB) were unable to detect dissociation events for incidence energies $\mathrm{E_{i}}$ up to 1.3 eV (i.e.\ almost twice the value of the estimated energy barrier), and only
molecular adsorption was observed at low surface temperatures, $\mathrm{T_{s}} \sim 90$~K\cite{Funk2006}.

Recent SMB studies by Singh and Shirhatti have shed new light on these apparently contradicting results \cite{Singh2024a}. 
They measured initial dissociative sticking probabilities, S$_0$, varying from $\sim 3\times 10^{-4}$ for $\mathrm{E_{i}}$=0.65 eV to $\sim 2\times 10^{-2}$ for $\mathrm{E_{i}}$=1.59 eV. Thus, in spite of 
observing dissociation events at relatively low impact energies, these results are not in contradiction with those of Funk \textit{et al.}
who would have not observed signs of CO$_2$ dissociation due to the detection limit of their experiments ($\sim 0.03$).
In addition, based on their results and extrapolations to higher impact energies through considerations based on a low-dimensional description of the CO$_2$ dissociation dynamics, Singh and Shirhatti estimated the energy barrier for CO$_2$ dissociation to be $\mathrm{E_{b}} \gtrsim$ 2 eV \cite{Singh2024a}, in strong contrast with density functional theory (DFT) results that predict an energy barrier $\mathrm{E_{b}}$=0.64 eV \cite{Wang2003,Yang2020}. 
Motivated by the latter work, Yin and Guo performed full dimensional quasi-classical trajectory (QCT) calculations using an artificial neural network potential energy surface (ANN-PES) based on DFT calculations (with the optPBE-vdW functional \cite{Klimes2010}) and obtained dissociative adsorption probabilities, $\mathrm{P_{diss}}$, smaller than 1.5$\times$10$^{-2}$ even for an impact energy more than three times higher than the lowest energy barrier for dissociation of their ANN-PES, $\mathrm{E_{b}}$=0.63 eV. 
Thus, they showed that the low reactivity of Cu(110) found in experiments for CO$_2$ is not only due to a relatively 
large energy barrier for dissociation but also to the tight character of the transition state and the complexity of the high-dimensional molecule-surface CO$_2$/Cu(110) interaction dynamics\cite{Yin2024}.

In a recent attempt to extend the impact energy range of their previous
SMB experiments, Singh and Shirhatti performed new measurements
for $\mathrm{E_{i}}$ up to 4.6 eV \cite{Singh2024b}.
In contrast with the rapid increase of $\mathrm{P_{diss}}$($\mathrm{E_{i}}$) observed at lower energies, for $\mathrm{E_{i}} >$ 3 eV $\mathrm{P_{diss}}$ tends to level off at a value $\sim 4.1 \times 10^{-2}$, increasing only by a factor 1.5 in the wide 2.0--4.6 eV $\mathrm{E_{i}}$-range. 
Furthermore, they also found that the O atom saturation coverage
obtained for large exposures of CO$_2$ molecules with $\mathrm{E_{i}} \gtrsim$ 
3 eV, is larger than for lower impact energy molecules (0.66 ML vs.\ 0.5 ML). 
This might indicate that for molecules with $\mathrm{E_{i}}$ higher than 3 eV, there are final states of the dissociation products  not accessible at lower impact energies,
but this has not yet been elucidated experimentally.  

In this work we revisit the dynamics of CO$_2$ dissociation on Cu(110)
through QCT calculations based on an ANN-PES parameterized from DFT total energies obtained using the vdW-DF2 exchange-correlation functional \cite{Lee2010} used previously with success for CO/Cu(110)\cite{Gonzalez2023,Gonzalez2025}.
We focus our study on the dynamics of the CO$_2$ molecules
with impact energies in the whole $\mathrm{E_{i}}$-range investigated in Refs. \citenum{Singh2024a,Singh2024b}.
We analyze in detail all the possible final states of the molecule, for molecular and dissociative adsorption.
In particular, we investigate and quantify to what extent the impact energy and surface temperature influence the final state of the dissociation products, including the generation of surface defects.

\section{Methodology}

\subsection{DFT Calculations} \label{DFT}

Total energies for the CO$_{2}$/Cu(110) system were computed using the nonlocal van der Waals density functional vdW-DF2, originally proposed by Lee \emph{et al.} \cite{Lee2010}, to account for dispersion interactions.
We have used the Vienna ab initio simulation package (VASP) \cite{Kresse1993a,Kresse1994a,Kresse1994b,Kresse1996a,Kresse1996b,Kresse1999}, which employs a plane-wave basis set to represent electronic wavefunctions and accounts for electron-ion interactions through the projector augmented-wave (PAW) method\cite{Blochl1994}.
The cutoff energy for plane wave expansions was set at 450 eV, and the Methfessel–Paxton scheme \cite{Methfessel1989} was used with a smearing width of 0.4 eV. The Cu(110) surface was modeled as a five-layer slab substrate with a vacuum of 13.6 {\AA} in the direction normal to the surface. 
The Cu atoms in the two bottom layers were kept fixed with the ideal inter-layer distance between consecutive 110 planes in bulk, $a_{Cu}\sqrt{2}/4$, with $a_{Cu}$ being the obtained optimum theoretical lattice constant, $a_{Cu}$=3.754 \AA. In contrast, the positions of the Cu atoms in the three topmost layers of the Cu(110) slab were optimized to account for the relaxation of the clean surface and also in geometry optimizations of adsorbate/surface structures.
We performed DFT calculations for (3$\times$2) and (3$\times$3) supercells   
using 7$\times$7$\times$1 and 7$\times$5$\times$1 Monkhorst–Pack {\bf k}-point grids\cite{Monkhorst1976} respectively.
In general, the DFT calculations were spin restricted except for those configurations with the O atom far from the surface for which we performed spin-polarized calculations.
All these settings of the DFT calculations are consistent with those used previously in Ref. \citenum{Gonzalez2023} to deal with CO/Cu(110). 
 
 \begin{figure}
 	\centering
 	\includegraphics[width=0.5\linewidth]{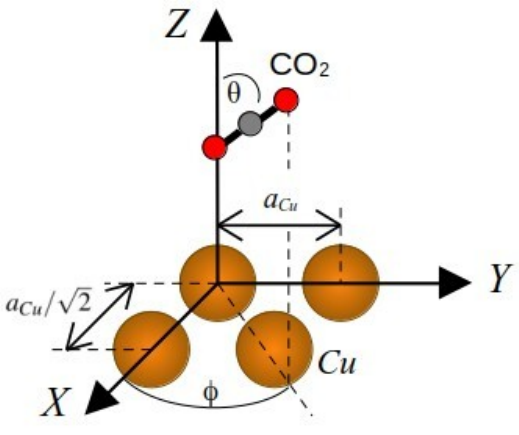}
 	\caption{Schematic representation of the CO$_2$ molecule in its equilibrium configuration far from the surface, and  coordinate system used throughout this work. $\mathrm{Z}$=0 corresponds to the plane containing the outermost-layer Cu atoms, in the lowest-energy clean surface structure. High symmetry sites: top ($\mathrm{X}$=0,$\mathrm{Y}$=0);
 	short-bridge ($\mathrm{X}$=a$_{Cu}/(2\sqrt{2})$,$\mathrm{Y}$=0); long-bridge ($\mathrm{X}$=0, $\mathrm{Y}$=a$_{Cu}/2$); hollow ($\mathrm{X}$=a$_{Cu}/(2\sqrt{2})$,$\mathrm{Y}$=a$_{Cu}/2$).}   
 	\label{fig:CO2_coordinates}
 \end{figure}
 
All geometry optimizations were performed until the forces acting on all mobile atoms were smaller than 0.1 eV/\AA.
The reported optimum geometry of CO$_{2}$ in vacuum was actually obtained
in calculations with the molecule in the middle of the $\sim 13.6$~\AA\
vacuum space left  between the slab and its closest periodic images.
The equilibrium C--O bond lengths found for the optimum linear configuration of CO$_2$ are both 1.177 \AA.
The CO$_2$ molecule in its equilibrium configuration far from the surface
is shown in Fig.\ \ref{fig:CO2_coordinates} where we also illustrate 
the system coordinate used along this work. 
The minimum energy pathways (MEP) for all the reactive processes we have investigated were computed using the nudged elastic band (NEB) \cite{Mills1994,Mills1995,Jonsson1998} and dimer \cite{Henkelman1999} methods.
The character of true saddle point of the PES for all the transition states
(TS) reported in this work was confirmed by checking the existence of only one negative eigenvalue of the Hessian matrix of the DFT-PES.

\subsection{The ANN-PES} \label{ANN-PES}  

From the DFT total energies for a large set of configurations of the CO$_2$/Cu(110) system, we developed a PES using the atomistic neural network (ANN) method  implemented in the atomic energy network (\texttt{{\ae}net}) code \cite{Artrith2016, Artrith2017, Cooper2020}.
Using this method, the total energy of the system, $\mathrm{E_{ANN}}$, is expressed as the sum of a chemical-environment-dependent energy of each atom in the supercell (considering periodic boundary conditions):
\begin{equation}
	E_{ANN} =  
	\sum_{\alpha=1}^{N_{type}} \sum_{i=1}^{N_{\alpha}} E_i^{(\alpha)},
	\label{eq1}
\end{equation}
where $\mathrm{E_{i}^{(\alpha)}}$ is the energy of the $i^{th}$ atom of species $\alpha$ ($\alpha$=Cu, C, O), $\mathrm{N_{type}}$ is the number of atomic species, and $\mathrm{N_{\alpha}}$ is the number of atoms of species $\alpha$.
The chemical environment of each atom was described using the radial (G\textsuperscript{2}) and angular (G\textsuperscript{4}) descriptors introduced by Behler and Parrinello (see Ref. \citenum{Behler2017} and references therein) 
%with a cutoff radius of $R_c = 7.0$~{\AA}, 
as we have successfully used for CO/Cu(110)\cite{Gonzalez2023}.
For the atoms of the three species, $\mathrm{E_{i}^{(\alpha)}}$ was computed using a feed-forward neural network with two hidden layers and 10 neurons per layer, and the hyperbolic tangent as activation function.

For the G\textsuperscript{2} and G\textsuperscript{4} symmetry functions involving pairs and triads of species necessary to describe the system CO/Cu(110), here we have used the same parameters as in Ref. \citenum{Gonzalez2023}. 
For CO$_2$/Cu(110) one extra O-O radial function, and four extra angular functions
(Cu-O-O, C-O-O, O-Cu-O, and O-C-O) are needed due to the extra O atom.
For the O-O radial function, we have used: $\eta$=0.05, 0.23, 1.07, 2.0 with R$_{\mathrm{s}}$=0.0 and R$_{\mathrm{c}}$=7.0 \AA, whereas for the four extra angular symmetry functions
we have used: $\eta=0.05$, $\lambda=\{-1,1\}$, $\zeta=\{1,4\}$ and R$_{\mathrm{c}}$=7.0 \AA.

In order to optimize the ANN-PES for CO$_2$/Cu(110) we have used an active learning
approach. We have started with an initial database of configurations whose DFT total energies are used to minimize the root mean square error (RMSE) of the energies predicted by five ANN-PESs:

\begin{equation}
	RMSE= \left [  \frac{1}{N_{conf}} \sum_{j=1}^{N_{conf}} (E_{DFT,j}-E_{ANN,j})^2 \right ]^{1/2},
	\label{RMSE}
\end{equation}
where $\mathrm{N_{conf}}$ is the number of configurations in the database, and E$_{\mathrm{DFT,j}}$
and E$_{\mathrm{ANN,j}}$ are the DFT total energy and the ANN-PES value for the $j^{th}$
configuration respectively.
Actually, only 90\% of the configurations in the database have been used for training because 10\% of the configurations were selected (randomly) for testing, in order to detect signs of overfitting \cite{Behler2017}.
The five ANN-PESs differ with respect to each other only due to the different initial values of the weights and bias of the ANNs which are chosen randomly. 
Among the five initial ANN-PES, we selected the one with the lowest RMSE for the training and testing sub-sets of configurations, and used it to calculate MEPs for various process of interest (using the climbing image NEB, CI-NEB,  method
\cite{Jonsson1998,Henkelman2000a,Henkelman2000b,Smidstrup2014,Kolsbjerg2016,Makri2019} as implemented in the ASE package\cite{Bahn2002,Larsen2017}), and to perform quasi-classical trajectory (QCT) calculations for various impact energies of the impinging CO$_2$ molecules on Cu(110).
From these calculations, we selected configurations to evaluate the energy using the five ANN-PESs. 
For each configuration we compute the maximum discrepancy between the values predicted by the five ANN-PESs, and for those whose largest discrepancy exceeded 0.05 eV, we performed a DFT total energy evaluation and incorporated the configuration to the database. 

\begin{figure}
	\centering
	\includegraphics[width=0.35\linewidth]{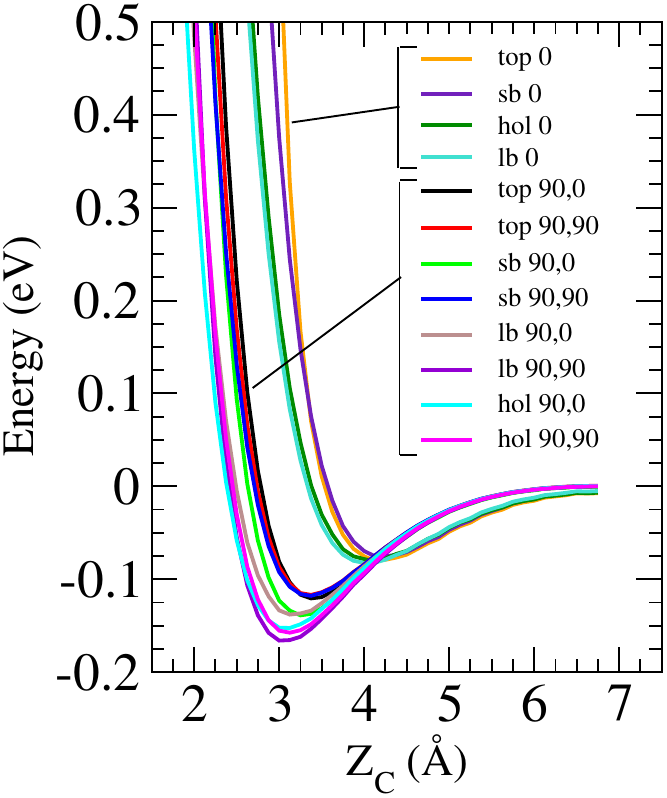}
	\caption{DFT-vdW-DF2 total energy of CO$_2$ in its equilibrium geometry in vacuum, as a function of the molecule surface distance measured by the $\mathrm{Z}$ coordinate of the C atom above the outermost-layer of surface atoms which are kept fixed in their equilibrium positions. Each line corresponds to a different surface high symmetry site: top, short-bridge (sb), long-bridge (lb), and hollow (hol) and/or a different orientation as indicated in the inset by the values of the polar and azimuthal angles of the molecule. For instance, hol 90,0 (lb 0) indicates that the center of mass of the molecule is on a hollow (long-bridge) site and the polar and azimuthal angles of (the polar angle of) the molecular axis containing the two C--O bonds are 90 deg and 0 deg respectively (is 0 deg).}   
	\label{fig:vdW:pot}
\end{figure}

The initial database for CO$_2$/Cu(110) included the full database of 7\,803 configurations used to parameterize the ANN-PES for CO/Cu(110) employed in Refs. \citenum{Gonzalez2023,Gonzalez2025}, plus extra configurations of
CO$_2$/Cu(110), (CO+O)/Cu(110) (dissociated states) and also a few for Cu(110) 
using two supercells: (3$\times$2) and (3$\times$3). 
The great majority of these extra configurations correspond to:
\begin{itemize}
	\item[(i)] CO$_2$ in its equilibrium geometry, and	vibrating in its different normal modes (asymmetric stretching, AS, symmetric stretching, SS, and bending, B) far from the surface,
	\item[(ii)] CO$_2$ in its equilibrium geometry on the four highest symmetry surface sites of Cu(110) (i.e.\ top, short-bridge, long-bridge and hollow), 
	parallel and perpendicular to the surface at different heights of the center of mass of the molecule above the outermost-layer Cu atoms which are kept 
	fixed in their equilibrium positions. The results of these calculations are shown in Fig.\ \ref{fig:vdW:pot},
	\item[(iii)] idem to (ii) but with the C--O bond lengths changed with respect to the equilibrium value by $\pm$ 0.1, 0.2 \AA\ (consistent with vibrations in the AS and the SS mode), and with angles between the two C--O bonds of 140, 145, 150, 155, 160, 165, 170, 175, and 180 deg,
	\item[(iv)] CO$_2$/Cu(110) evaluated using DFT calculations while looking for local minima of the PES corresponding to possible final states of CO and O after dissociation (close and far from each other), the optimum geometry of physisorbed states, and the bent molecularly chemisorbed state reported in the literature\cite{Yin2024}, 
	\item[(v)] images of DFT-based-NEB calculations explored while looking for MEPs
	and transition states (TS) from a physisorbed to the chemisorbed bent state (TS1),
	and from a physisorbed to the dissociated state (TS2),	and
	\item[(vi)] (O+CO)/Cu(110) for CO fixed on top of Cu atom
	in its optimum chemisorption configuration, and the O atom on the high-symmetry surface sites (near and far from CO), with its $\mathrm{Z}$ coordinate (measured with respect to the outermost-layer Cu atoms), $\mathrm{Z_{0}}$, varying from 0 to 6.8 \AA\ and 
	with O in the positions visited by the DFT geometry optimization performed looking for local minima of the PES.
\end{itemize}

During the iterative active learning method, we applied 
a descriptor-based filter to the database to avoid redundant-information (too similar) configurations (characterized by discrepancies between all the descriptors $<$0.5\% and energy differences $<$5 meV).
The active-learning loop was repeated ten times until obtaining stable predictions with the resulting ANN-PES, in particular for the geometry and energy of relevant intermediate, final (dissociated) and transition states, as well as for the dissociative adsorption probability ($\mathrm{P_{diss}}$) at different impact energies. 
At the end, the final database 
consists of  \,81\,657 configurations with energies computed using DFT calculations. 

\subsection{QCT calculations}

QCT calculations were performed for initially non-rotating CO$_{2}$ molecules impinging on Cu(110) at normal incidence. 
We employed the gas-surface reaction dynamics (GSRD) code developed in our group, which has been previously used to investigate scattering and adsorption for various molecule/surface systems  
using Tersoff-like reactive force fields (RFF)\cite{Shen2014,Lozano2015,Seminara2019,Moiraghi2020,Peludhero2021} and more recently also  with ANN-PESs \cite{Rivero2021,Gonzalez2023}.
This code integrates the equation of motion of the system using the direct Beeman method\cite{Schofield1973}, a variant of the Verlet integration algorithm \cite{Verlet1967,Verlet1968}.
As mentioned in the previous section, these calculations were performed not only to compute observables in production stage but also during the iterative optimization of the ANN-PES.

Once the final version of the ANN-PES was obtained, for each initial condition 
corresponding to a molecular impact energy $\mathrm{E_{i}}$ ranging between 0.01 eV and 4.5 eV and a surface temperature $\mathrm{T_{s}}$ between 50 K and 400 K, 
we have integrated 10\,000 trajectories with a maximum propagation time of $\mathrm{t_{sup}}$=100 ps and a time step of 0.5 fs.
For each trajectory, the initial positions and velocities of the Cu atoms
were randomly selected from a large set of {\em snapshots} generated previously
during a long NVT simulation for the clean surface at the $\mathrm{T_{s}}$ value of interest, controlled through a Berendsen thermostat \cite{Berendsen1984}.

In order to simulate initially non-rotating CO$_2$ molecules in the vibrational ground state (GS) and in vibrationally excited states of the asymmetric stretching (AS), symmetric stretching (SS), and bend (B) modes, 
the initial coordinates and velocities of their C and O atoms (with respect to the molecular CM) were selected following the standard procedure detailed in reference \cite{Sewell1997}, and selecting randomly the molecular orientation.
Then, the CO$_{2}$ center of mass (CM) was initially positioned at a distance of 8.5 {\AA} above the topmost-layer of the Cu(110) surface ($\mathrm{Z_{CM}}$=8.5 {\AA}) and the lateral coordinates, $\mathrm{X_{CM}}$ and $\mathrm{Y_{CM}}$, were uniformly sampled randomly within the whole simulation supercell.
Finally, the desired initial velocity of the molecular CM 
directed perpendicularly towards the surface for normal incidence conditions was added to all the atoms of the molecule.

In the QCT calculations we have considered as reflected to those molecules for which,
after interacting with the surface, the $\mathrm{Z}$ coordinate of the CM reaches a value
larger than in the initial state, i.e.\ $\mathrm{Z_{CM}} >$ 8.5 \AA\ and the CM momentum vector points toward the vacuum. 
Whenever this condition is fulfilled, we stopped the integration of the trajectory and the time is considered as its {\em reflection time}. 
On the other hand, we have considered that a dissociative adsorption event has taken place whenever one of the two C--O bonds reaches a length equal to 1.8 \AA\ 
and we define that time as {\em dissociation time}. However, in this case
we have continued the integration of the trajectories up to 10 ps in order to  characterize better the {\em final state} of the dissociation products (allowing molecule-surface energy exchange and dissipation during several picoseconds after the dissociation time).
Finally, for those trajectories for which none of the two conditions mentioned above is fulfilled at a time $\mathrm{t}=\mathrm{t_{sup}}$=100 ps, we consider that the molecule has been molecularly adsorbed.
Accordingly, we define reflection, dissociative adsorption, and molecular adsorption probabilities ($\mathrm{P_{ref}}$, $\mathrm{P_{diss}}$, and $\mathrm{P_{molads}}$, respectively) as the ratio of the number of trajectories that fulfill the corresponding final condition and the total number of trajectories integrated for a given initial condition defined by the impact energy and vibrational state of the molecule and the surface temperature.   

\section{Results and discussions} \label{sec:res}

\subsection{ANN-PES validation}

\begin{figure}%[H]
	\centering
	\includegraphics[width=0.42\linewidth]{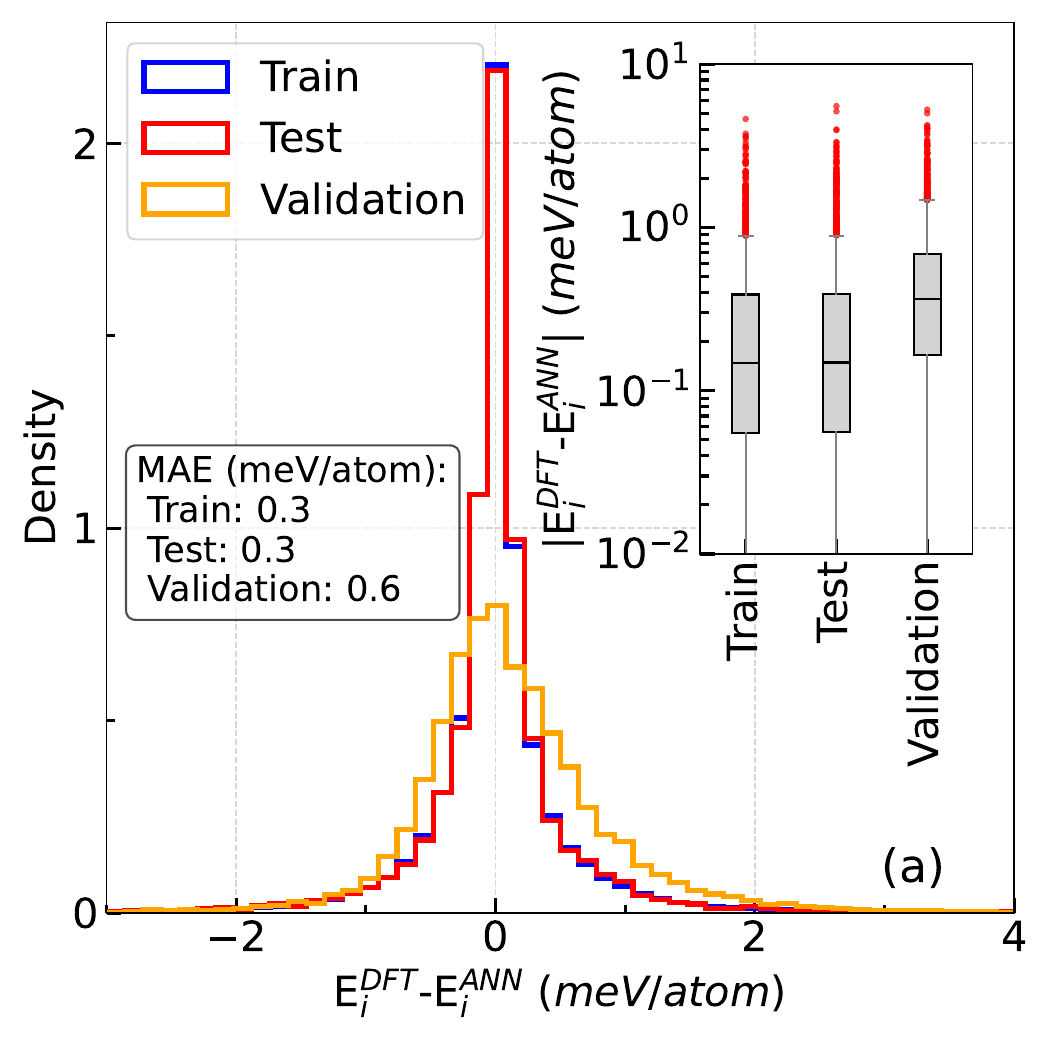}
%	\hfill
	\includegraphics[width=0.42\linewidth]{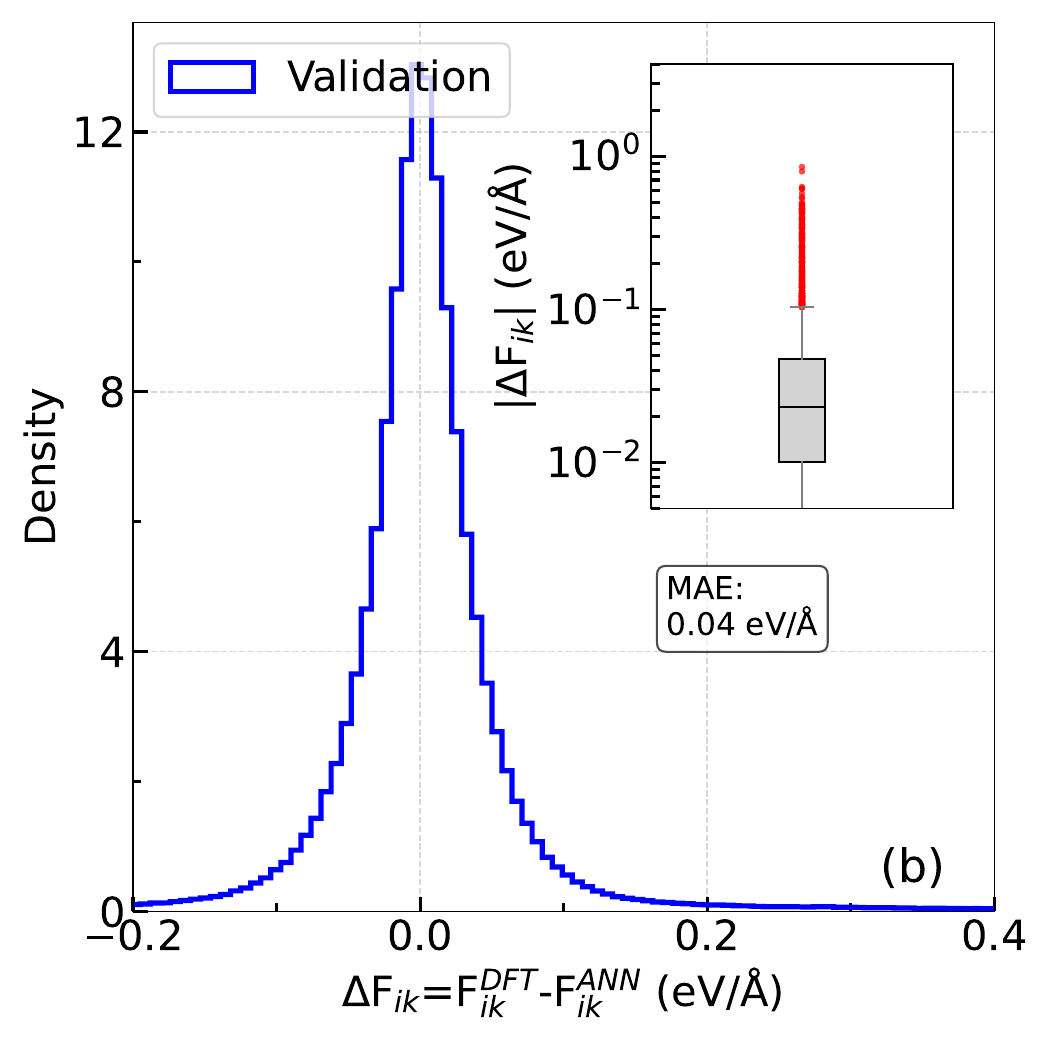}
	\caption{(a) Histogram of energy differences (E$_\text{DFT}$ - E$_\text{NN}$) for the training, test, and validation sets. (b) Force prediction errors for the validation set.}
	\label{fig:histogram_edft_enn}
\end{figure}

As already mentioned in section \ref{ANN-PES}, 90\% of the configurations of the total database were used for training (training set) and the other 10\% were used for testing (testing set).
In addition, during the training process we have selected a set of 12\, 675 configurations from the QCT calculations, whose total energy was evaluated using DFT but they were not incorporated 
to the database, in order to use them for validation (validation set).

Fig.~\ref{fig:histogram_edft_enn}(a) shows the histograms of energy errors per atom (E$_{\text{DFT}}$-E$_{\text{ANN}}$) for the training, testing and validation sets of configurations. It can be seen that most of the values are in the interval from -2 to 2 meV/atom.
Within the figure we also report the mean absolute error (MAE) per atom for the energy,
\begin{equation}
	\text{MAE}_{\mathrm{E}}=\frac{1}{N_{conf}} \sum_{j=1}^{N_{conf}} \frac{1}{N_{atom,\,j}}\left| E_{DFT,j}-E_{ANN,j} \right|  \;,
\end{equation}
with $\mathrm{N_{conf}}$ being the number of configurations in the corresponding set, and N$_{\mathrm{atom,\,j}}$ the number of atoms in the $j^{th}$ configuration. 
In addition, in the inset we show the boxplots (as implemented in matplotlib\cite{Hunter2007}) of the absolute errors 
for the three sets of configurations.
They show that 50\% (75\%) of the configurations in the training, testing and validation sets have absolute errors smaller than  0.15, 0.15 and 0.37 meV/atom (0.39, 0.39, and 0.69 meV/atom) respectively. These similar and small absolute errors
obtained for the three sets of data is a strong indication that our model is not overfitted.
There are some outlier configurations (represented by red symbols) with absolute errors up to $\sim 8$~meV/atom, but they represent only 
the $\sim 6.6$\% of the configurations and it is not expected 
they could produce significant artifacts in the dynamics.

To quantify the errors in the forces predicted with the ANN-PES with respect to the DFT values, we have also computed the MAE for the forces
using the expression:
\begin{equation}
	\text{MAE}_{\mathrm{F}}=\frac{1}{N_{conf}} \sum_{j=1}^{N_{conf}}\,\,\sum_{k=1}^{N_{atom,\,j}}  
	\frac{1}{N_{atom,\,j}}\,\,\sum_{\alpha} \left| F_{DFT,jk\alpha}-F_{ANN,jk\alpha} \right|  \;,
\end{equation}
where $\mathrm{F_{ANN,\,jk\alpha}}$ ($\mathrm{F_{DFT,\,jk\alpha}}$) is the $\alpha$-component
($\alpha=x,y,z$) of the ANN-PES (DFT) force acting on the $k^{th}$ atom of the $j^{th}$ configuration.
For the validation set (characterized by larger errors than the training and testing ones) we obtained MAE$_{\mathrm{F}}$=0.04 eV/\AA. 
Fig.~\ref{fig:histogram_edft_enn}(b) shows the histogram for the errors in the forces $\Delta \mathrm{F_{jk}}$ for the validation set.
It is observed that the great majority of them lie between -0.2 and 0.2 eV/\AA.
In the inset we also show the boxplot for the validation set which 
exhibits a relatively small set of outliers with errors from 0.1 to 1 eV/\AA. 

It is important to mention that during the last steps of the iterative active learning procedure, the values of $\mathrm{P_{diss}}$ for various impact energies differ by no more than a few percent with respect to the values predicted with the final ANN-PES which indicates
a good convergence of our procedure with respect to the observable of most interest for this work. 
In addition, we have found that using the present ANN-PES developed for CO$_2$/Cu(110), the molecular adsorption probabilities for CO/Cu(110) under normal incidence differ by less than 2\% with respect to the values reported in Ref. \citenum{Gonzalez2023}.

To further validate the accuracy of the ANN-PES, 
we use it for a systematic exploration of the CO$_2$/Cu(110)
potential energy landscape including determination of local minima
and minimum energy pathways for molecular and dissociative adsorption, as well as diffusion of the dissociation products.
The most relevant configurations found in these calculations were then evaluated using DFT and we have obtained discrepancies not larger than 30 meV.

\subsection{Characterization of the CO$_2$/Cu(110) ANN-PES} \label{sec:valANN-PES:nebs}

\begin{figure}
	\centering
	\includegraphics[width=1\linewidth]{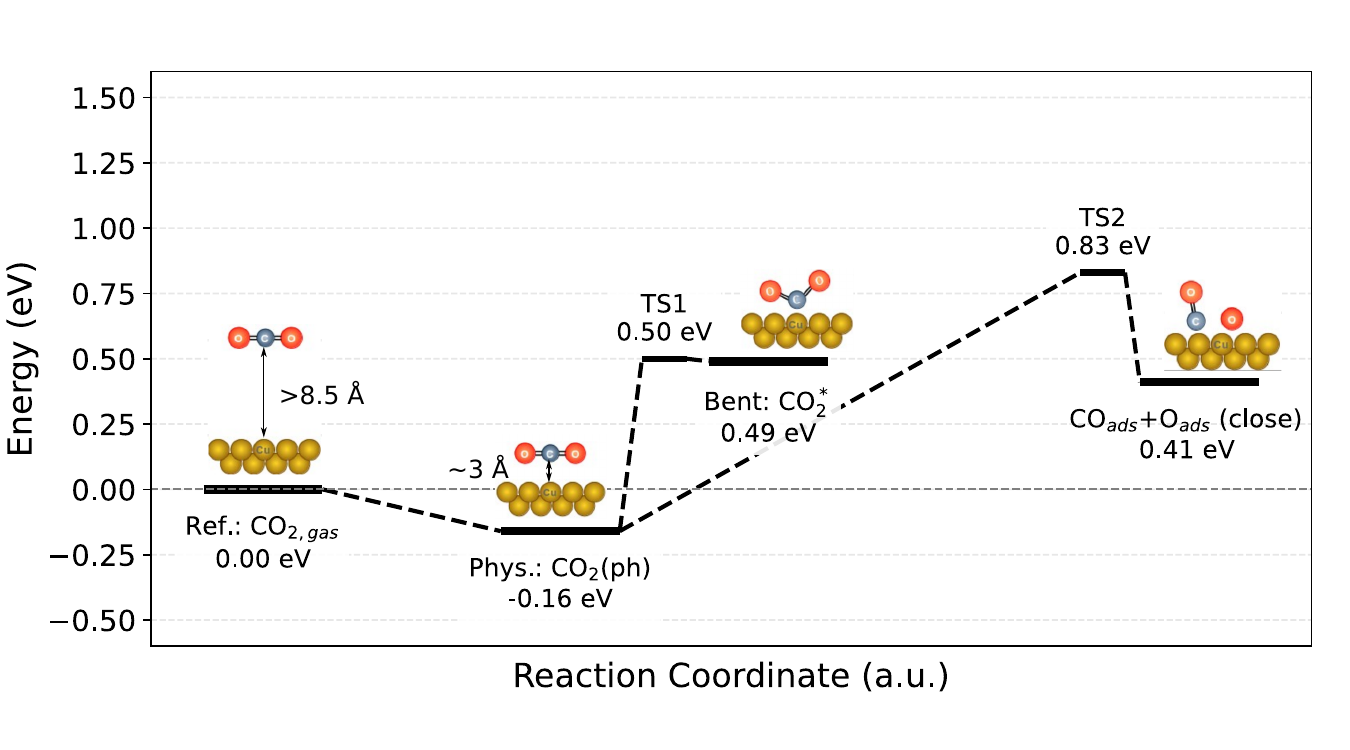}
	\caption{Energetics of CO$_2$ physisorption, molecular chemisorption and dissociation on Cu(110), including the involved local minima and transition states found along minimum energy pathways connecting them.}
	\label{fig:energy_paths}
\end{figure}

By exploring the properties of the ANN-PES of CO$_2$/Cu(110)
relevant for dissociative adsorption, we have identified four local minima corresponding to: 
(i) both the CO$_2$ molecule and the surface in their equilibrium configurations when they are from each other, $\mathrm{CO_{2,gas}}$; 
(ii) the physisorbed state of CO$_2$, CO$_2$(ph); 
(iii) a chemisorbed bent configuration of CO$_2$, CO$_2^*$; and 
(iv) the dissociated state, $\mathrm{CO_{ads}}$+$\mathrm{O_{ads}}$, with the two
dissociation products close to each other (within the same surface unit cell). 
The energy of these four local minima (taking $\mathrm{CO_{2,gas}}$ as the reference zero-energy configuration) and the transition states along the minimum energy pathways (MEP) connecting them are shown in Fig.\ \ref{fig:energy_paths}.

Physisorption is a non-activated process and there is no energy barrier
along the MEP connecting the CO$_2$(g) and CO$_2$(ph) states.
CO$_2$(ph) corresponds to a weakly bound molecular configuration due
to the van der Waals interaction with Cu(110).
In this configuration the molecule is located parallel to the surface 
(with its equilibrium geometry barely changed with respect to vacuum) with the $\mathrm{Z}$ coordinate of its center of mass, $\mathrm{Z_{CM}}\sim 3$~\AA.
The energy of this physisorbed state is -0.16 eV which is consistent with the value previously reported using the same vdW-DF2 functional \cite{Muttaqien2017} but corresponds to a desorption energy slightly smaller than the value extracted from temperature programmed desorption (TPD) experiments\cite{Ernst1999}.
This state is the global minimum of the CO$_2$/Cu(110) PES.
Fig.\ \ref{fig:vdW:pot} indicates that physisorbed molecules can
diffuse over the surface by overcoming an energy barrier $\sim 0.05$~eV.

The dissociated state with CO and O adsorbed within the same surface unit cell (with CO on top of a Cu atom and O occupying 
a pseudo three-fold adsorption site forming bonds with two nearest neighbor outermost-layer Cu atoms and a Cu atom in the second surface layer) has the energy +0.41 eV.
Finally, we have also found a bent chemisorbed state that exhibits
a tridentate molecular configuration with each atom of the molecule
bound to a different Cu atom of the surface unit cell and the angle O--C--O being $\sim 123$~deg.
Such a bent configuration is usually associated with a partial charge transfer from the surface to the antibonding $\pi^*$ orbitals of CO$_2$, thereby facilitating its activation toward dissociation or reduction\cite{Higham2020}.
The obtained optimum geometry of this bent-CO$_2^*$ state is very similar to the one reported previously characterized by a O--C--O angle
between 120 deg and 140 deg \cite{Etim2021, Yin2024}.
However, its energy (i.e.\ 0.49 eV) is significantly higher than
the value obtained by Yin and Guo (i.e.\ 0.06 eV) which is very likely 
due to the use of different XC functionals (vdW-DF2 used here vs.\
optPBE-vdW used in Ref. \citenum{Yin2024}).
 
\begin{figure}
	\centering
	\includegraphics[width=1\linewidth]{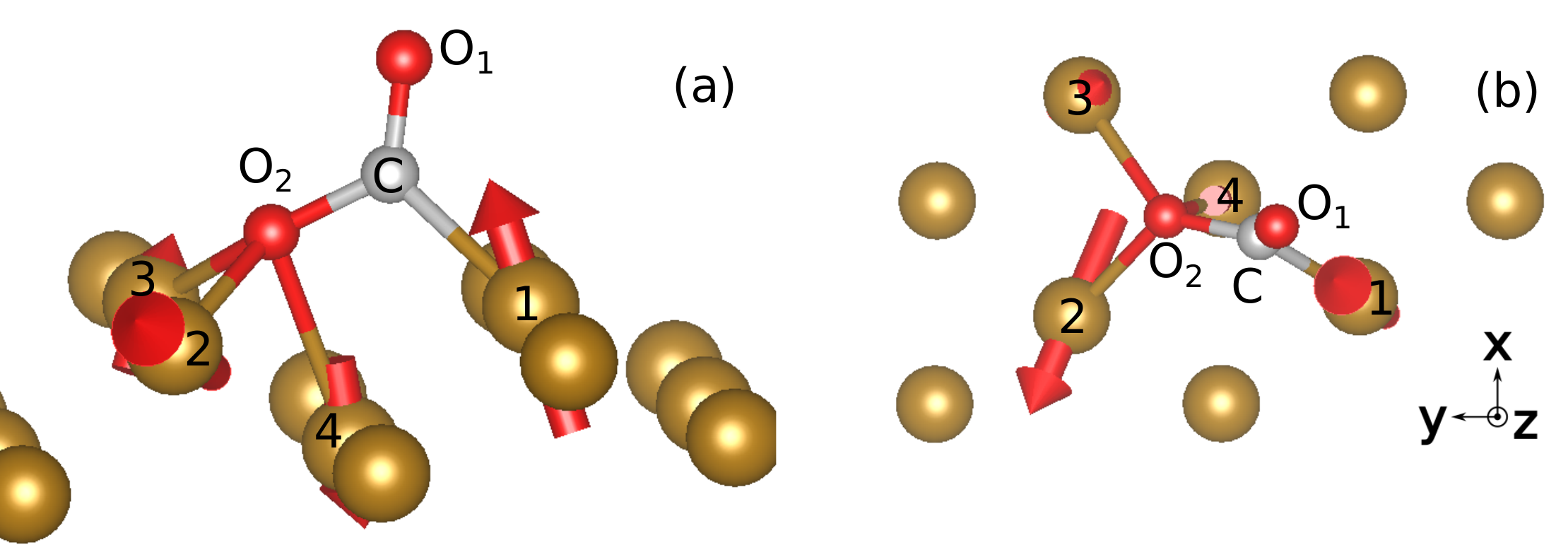}
	\caption{(a) Top view, and (b) side view of the transition state TS2 along the minimum energy pathway for CO$_2$ dissociation on Cu(110), connecting the physisorbed and the most stable dissociated state. Dark and light orange
	distinguish topmost-layer and second-layer Cu atoms respectively. Relevant distances/coordinates: d(C-O$_1$)=1.39 \AA, d(C-O$_2$)=1.21 \AA, $\mathrm{Z_{CM}}$=1.58 \AA, d(C-Cu$_1$)=2.05 \AA, d(O$_2$-Cu$_2$)=2.02 \AA, d(O$_2$-Cu$_3$)=2.13 \AA, d(O$_2$-Cu$_4$)=2.45 \AA. In (a) the red arrows represent the displacements of Cu atoms (larger than 0.1 \AA) with respect their positions in the reference configuration (see the text) of lengths:  0.23 \AA, 0.24 \AA, 0.33 \AA, and 0.15 \AA\ for the Cu atoms labeled as 1, 2, 3, and 4 respectively.}
	\label{fig:TS2_description}
\end{figure}

In Fig.\ \ref{fig:energy_paths} we also show the energies of two 
transition states along the MEPs connecting 
the physisorbed state with the bent-CO$_2^*$ state (TS1) and with the dissociated state (TS2). 
The energies of TS1 and TS2 (with respect to $\mathrm{CO_{2,gas}}$) are 0.50 eV and 0.83 eV respectively.
The optimum geometry (Fig.\ \ref{fig:TS2_description}) and the energy  of the TS2 structure (with respect to that of CO$_2$(ph)) obtained with our ANN-PES are very close to that of the transition state (also called TS2) reported by Yin and Guo\cite{Yin2024} using the optPBE-vdW functional (0.99 eV vs.\ 0.87 eV).
The energy barrier to escape from the bent-CO$_2^*$ state towards the CO$_2$(ph) state is only 0.01 eV. 
This entails a very small stability of the bent chemisorbed state.
The corresponding energy barrier reported by Yin and Guo is also  small (0.08 eV) but still larger than ours by 0.07 eV\cite{Yin2024}.
This small stability predicted for the CO$_2^*$ state is consistent with the results of the experimental investigations that have not been able to detect a bent chemisorbed state of CO$_2$ on the (110) face of copper\cite{Fu1992,Ernst1999}.

It must be mentioned that we have also tried to compute
a MEP connecting directly the bent-CO$_2^*$ state with the dissociated state (i.e.\ not passing through CO$_2$(ph)) but in spite of many attempts through NEB calculations with different settings, we did not succeed. All the obtained dissociation pathways starting from bent-CO$_2^*$, passed first by a more stable physisorbed state and the 
the energy barriers obtained were always very close to that of TS2 reported in Figs.\ \ref{fig:energy_paths} and \ref{fig:TS2_description}.

Before concluding this section it is important to emphasize that
the energy of all the relevant configurations displayed in Fig.\ \ref{fig:energy_paths} were evaluated with DFT calculations. The obtained discrepancies with respect to the ANN-PES values were always smaller than 0.03 eV. Moreover, DFT calculations confirmed the local minimum or saddle point character predicted by the ANN-PES for all
these configurations.
This provides strong support for the use of the ANN-PES for large-scale dynamical simulations of CO$_2$ scattering, adsorption and dissociation on Cu(110).

\subsection{QCT results} \label{sec:QCT_res}

\begin{figure}
	\centering
	\includegraphics[width=0.7\linewidth]{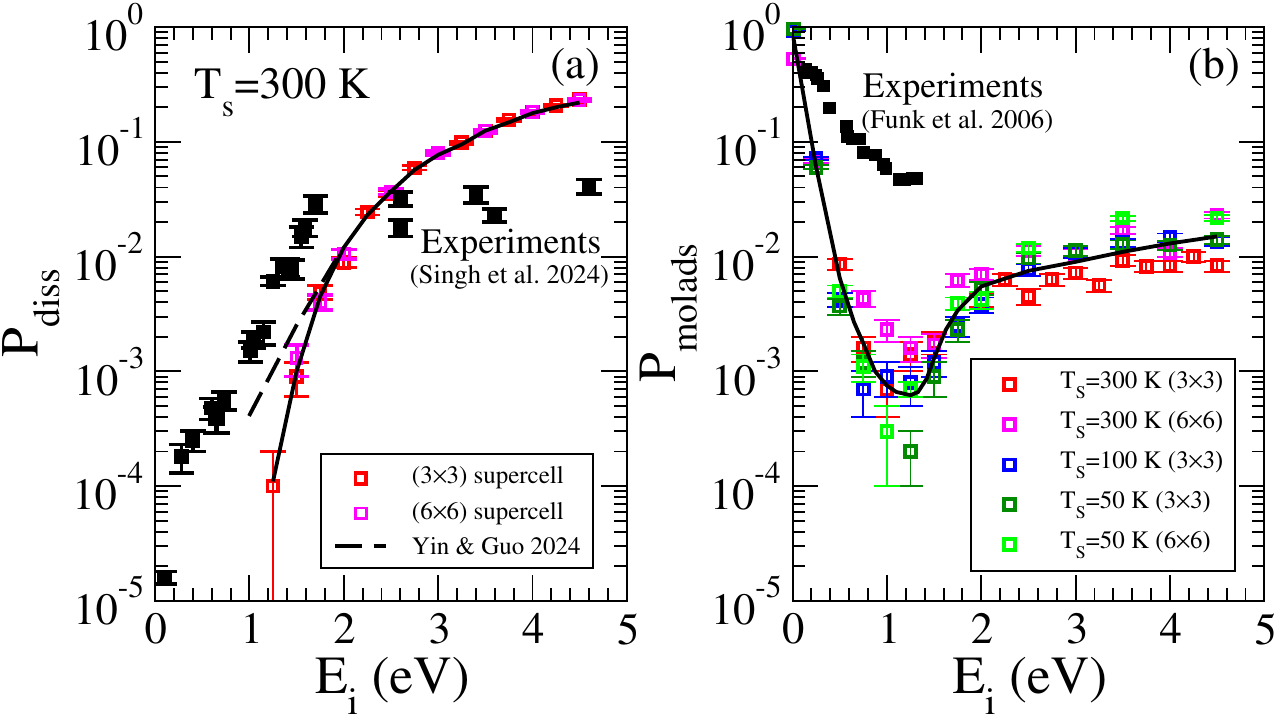}
	\caption{(a) $\mathrm{P_{diss}}$ as a function of the impact energy, $\mathrm{E_{i}}$, for normal incidence and $\mathrm{T_{s}}$=300 K. Full black squares, experimental data taken from Refs. \citenum{Singh2024a,Singh2024b}. Red open squares, QCT results for a ($3\times3$) supercell (present work), magenta open squares, QCT results for a ($6\times6$) supercell (present work). Dashed black line, QCT results taken from Ref. \citenum{Yin2024}. (b) $\mathrm{P_{molads}}$ as a function of the impact energy, $\mathrm{E_{i}}$, for normal incidence. Full black squares, experimental data taken from Ref. \citenum{Funk2006}. Open squares, QCT results (present work) for different surface temperatures and supercells: red, 300 K and ($3\times3$); magenta, 300 K and ($6\times6$); blue, 100 K and ($3\times3$); dark green, 50 K and ($3\times3$); light green, 50 K and ($6\times6$). The full black lines in both panels are plotted to guide the eyes.	}
	\label{fig:stick_GS_prob}
\end{figure}

In Fig.\ \ref{fig:stick_GS_prob}(a) we show the dissociative sticking probability, $\mathrm{P_{diss}}$, of CO$_2$ on Cu(110) as a function of the initial translational energy of the molecules, $\mathrm{E_{i}}$, under normal incidence, compared with available supersonic molecular beam experimental data for a surface temperature $\mathrm{T_{s}}$=300 K \cite{Singh2024a,Singh2024b}.
Our QCT results (for CO$_2$ initially in its roto-vibrational ground state) are also compared with those reported recently by Yin and Guo\cite{Yin2024}, obtained using a similar methodology.
We have considered two supercells: (3$\times$3) (as used by Yin and Guo\cite{Yin2024}) and (6$\times$6). 
Fig.\ \ref{fig:stick_GS_prob}(a) shows that the values of $\mathrm{P_{diss}}$ obtained with both supercells are very close to each other. 

As expected for activated dissociative adsorption,
$\mathrm{P_{diss}}$ shows a monotonic increasing $\mathrm{E_{i}}$-dependence in qualitative agreement with the supersonic molecular beam data.
However, compared with experiments our results are too low for $\mathrm{E_{i}} \lesssim$ 2 eV and too high for $\mathrm{E_{i}} \gtrsim $ 2 eV.
On the other hand, for $\mathrm{E_{i}} \lesssim $ 1.75 eV our results are also smaller than those of Yin and Guo, which is very likely due to the higher minimum energy barrier (MEB) for CO$_2$ dissociation of our ANN-PES: 0.83 eV vs.\
0.63 eV reported in Ref. \citenum{Yin2024}.
However, both theoretical results approach each other for higher values of $\mathrm{E_{i}}$, suggesting that for such high impact energies, $\mathrm{P_{diss}}$ becomes much less sensitive to the value of the MEB. 
This is not surprising since for impact energies well above the MEB,
dissociation can take place through an increasing number of reaction pathways, many of them not necessarily close to the MEP. 
 
Above ${\mathrm{E_{i}}}$ $\sim 2$~eV the experimental dissociative sticking probability becomes $\mathrm{E_{i}}$-insensitive in contrast with our theoretical results, though in this high-energy range the slope of the theoretical $\mathrm{P_{diss}}$($\mathrm{E_{i}}$) curve is much smaller than at low impact energies.
It is important to mention that above $\mathrm{E_{i}} \sim 3$~eV, in experiments an atomic oxygen saturation coverage higher than that found for lower impact energies was observed.
This might be an indication that the final states accessible for the dissociating CO$_2$ molecules on Cu(110) change from low to high impact energies.
This motivates our analysis of the time evolution of CO and O post-dissociation presented below.

For $\mathrm{E_{i}} \gtrsim$ 1 eV, most of the impinging CO$_2$ molecules that do not dissociate are reflected back to vacuum. However, there is a non-negligible
fraction of trajectories that remain (intact) near the surface 
even after a maximum integration time, $\mathrm{t_{sup}}$= 100 ps.
The corresponding probability of molecular adsorption 
is here referred to as $\mathrm{P_{molads}}$.
Fig.\ \ref{fig:stick_GS_prob}(b) shows that no matter the size of supercell
(3$\times$3) or (6$\times$6), and the surface temperature,   
$\mathrm{P_{molads}}$ presents a non-monotonic $\mathrm{E_{i}}$-dependence.
$\mathrm{P_{molads}}$ is very large for very low energies (varies between 0.5 and 1 depending on the considered value of $\mathrm{T_{s}}$), decreases with increasing $\mathrm{E_{i}}$
up to $\sim 1.25$~eV, and then increases from its minimum value ($\sim 10^{-3}$--10$^{-4}$) and tend to level off close to 10$^{-2}$ for $\mathrm{E_{i}}\gtrsim 2.5$~eV.
Thus, our calculations predict that molecular adsorption is the process with the highest probability at low impact energies ($\mathrm{E_{i}} \lesssim 0.1$~eV) and 
only for $\mathrm{E_{i}} \gtrsim$ 1.75 eV, $\mathrm{P_{molads}}$ becomes smaller than $\mathrm{P_{diss}}$.  %(we will analyze possible consequences of this in more detail below).
It might be argued that, $\mathrm{P_{molads}}$ values could change significantly with the size of the supercell employed in the calculations and $\mathrm{T_{s}}$, 
because both factors are expected to affect the energy transfer from the molecule to the surface, necessary to stabilize high energy molecules near the surface. However, this is not the case. 
The results obtained for both (3$\times$3) or (6$\times$6) supercells, and different $\mathrm{T_{s}}$ values between 50 K and 300 K, lie all rather close to a single $\mathrm{P_{molads}}$($\mathrm{E_{i}}$) curve (full black curve plotted to guide the eyes
in Fig.\ \ref{fig:stick_GS_prob}(b)).
For CO$_2$/Cu(110), Funk \textit{et al.} performed measurements of the 
trapping probability as a function of $\mathrm{E_{i}}$ at $\mathrm{T_{s}}$=90 K \cite{Funk2006}, i.e.\ close below the desorption temperature of CO$_2$ observed in temperature programmed experiments\cite{Ernst1999}.
The experimental molecular trapping probability is close to 0.4 at very low 
impact energies in reasonable agreement with our results
and also decreases (exponentially) with $\mathrm{E_{i}}$ but 
much less pronounced than the theoretical results.
For instance, the theoretical $\mathrm{P_{molads}}$ values for $\mathrm{E_{i}}$=1.25 eV are smaller than the experiments by two orders of magnitude. 
The reason for such a big discrepancy is unclear to us, but it is highly unlikely due to a possible slight underestimation of the physisorption well-depth in our ANN-PES.
In what follows, we will only present QCT results obtained using the (6$\times$6) supercell except otherwise stated. 

\begin{figure}
	\centering
	\includegraphics[width=0.6\linewidth]{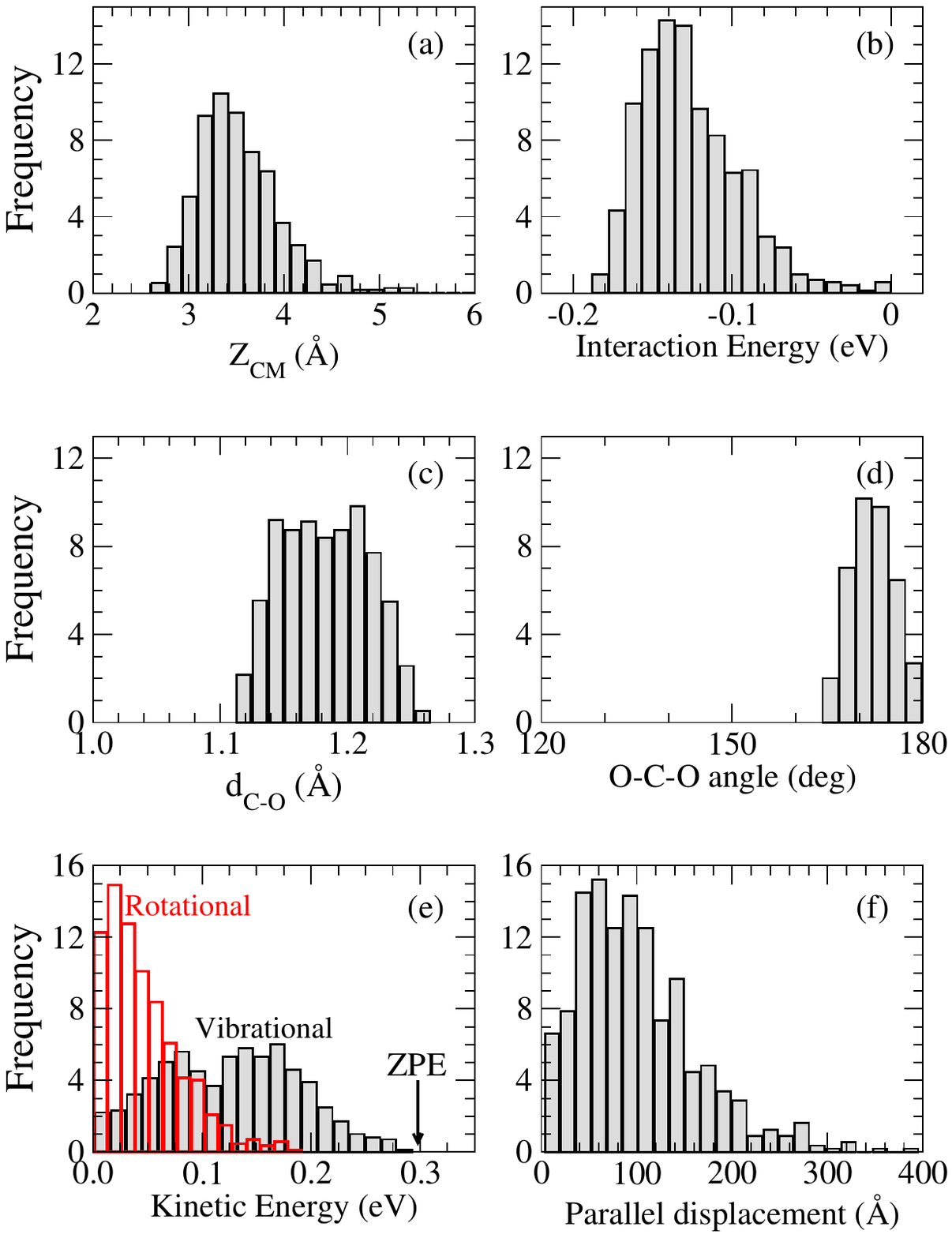}
	\caption{Characterization of non-dissociated CO$_{2}$ molecules adsorbed at the final integration time of $\mathrm{t_{f}}$=100 ps. Results correspond to calculations made in a (6$\times$6) supercell size, for normal incidence with initial kinetic energy $\mathrm{E_{i}}$=0.25 eV at a surface temperature $\mathrm{T_{s}}$=300 K. Distributions for: (a) final $\mathrm{Z}$ coordinate of the molecular center of mass, $\mathrm{Z_{CM}}$. (b) final interaction energy (see main text for definition). (c) final C-O bond length, $\mathrm{d_{C-O}}$. (d) final O-C-O bond angle. (e) final rotational (red histogram) and vibrational (grey histogram) kinetic energies. (f) Displacement of the molecular center of mass parallel to the surface.}   
	\label{fig:vdW:admol}
\end{figure}

The analysis of the final state of low energy molecules that remain trapped near the surface at $\mathrm{t_{sup}}$=100 ps shows that they are all with $\mathrm{Z_{CM}}$ between 2.6 \AA\ and 5.4 \AA\ (see Fig.\ \ref{fig:vdW:admol}(a), for $\mathrm{E_{i}}$=0.25 eV and $\mathrm{T_{s}}$=300 K). 
The molecule-surface interaction energy defined as the difference between the potential energy of the configurations $f$ and $f'$, $\Delta$E=E$_f$-E$_{f'}$ (with $f$ being final configuration and $f'$ is obtained from $f$ by only translating rigidly the CO$_2$ molecule 
to a position where the interaction with the surface vanishes) is shown in Fig.\ \ref{fig:vdW:admol}(b). The obtained small and negative $\Delta$E values are consistent with the weak attractive molecule-surface interaction near 
the physisorption well shown in Fig.\ \ref{fig:vdW:pot}. 
In addition, the distributions of the C-O distances (Fig.\ \ref{fig:vdW:admol}(c)) are close to the equilibrium value for CO$_2$ in vacuum, and that of the O-C-O angle between the two C-O bonds differ from 180 deg by less than 15 deg (Fig.\ \ref{fig:vdW:admol}(d)).
All these results demonstrate that molecules remaining trapped near the surface at $\mathrm{t_{max}}$=10 ps are physisorbed on the surface.
The distribution of kinetic energy corresponding to vibrational motion 
is consistent with the molecular ZPE=0.298 eV, 
whereas the distribution of
rotational kinetic energy shows signs of moderate rotational excitation (Fig.\
\ref{fig:vdW:admol}(e)).
Finally, it is interesting to mention that these physisorbed molecules 
that have been trapped near the surface as long as 100 ps have experienced
very large displacements parallel to the surface with respect to the initial aiming point as shown in Fig.\ \ref{fig:vdW:admol}(f).

Comparing the values of $\mathrm{P_{diss}}$ and $\mathrm{P_{molads}}$ shown in
 panels (a) and (b) of  Fig.\ \ref{fig:stick_GS_prob} respectively, it should be noted that $\mathrm{P_{molads}}$ is larger than $\mathrm{P_{diss}}$ by a factor 
 expected larger than 10$^5$ for $\mathrm{E_{i}} \lesssim$ 1 eV, that decreases to a value $\sim$ 1 for $\mathrm{E_{i}}$=1.5 eV, whereas
for $\mathrm{E_{i}} \gtrsim$ 1.75 eV $\mathrm{P_{diss}}$ becomes bigger than $\mathrm{P_{molads}}$.
Therefore, if in experiments a non-vanishing (even small) fraction of physisorbed molecules traveling large distances parallel to the surface (as shown in Fig.\ \ref{fig:vdW:admol}(f) for $\mathrm{E_{i}}$=0.25 eV) might encounter defect sites  (e.g.\ steps) offering lower-energy-barrier dissociation pathways, the observed dissociative adsorption probability at low impact energies might be significantly larger than the values predicted by our calculations for a  defect-free Cu(110) model surface. 
However, quantifying the role of defects certainly requires further modeling \cite{Jackson2020,Zhou2021} beyond the scope of this paper.  

\begin{figure}
	\centering     
	\includegraphics[width=0.8\linewidth,trim={1cm 6cm 0cm 0cm}]{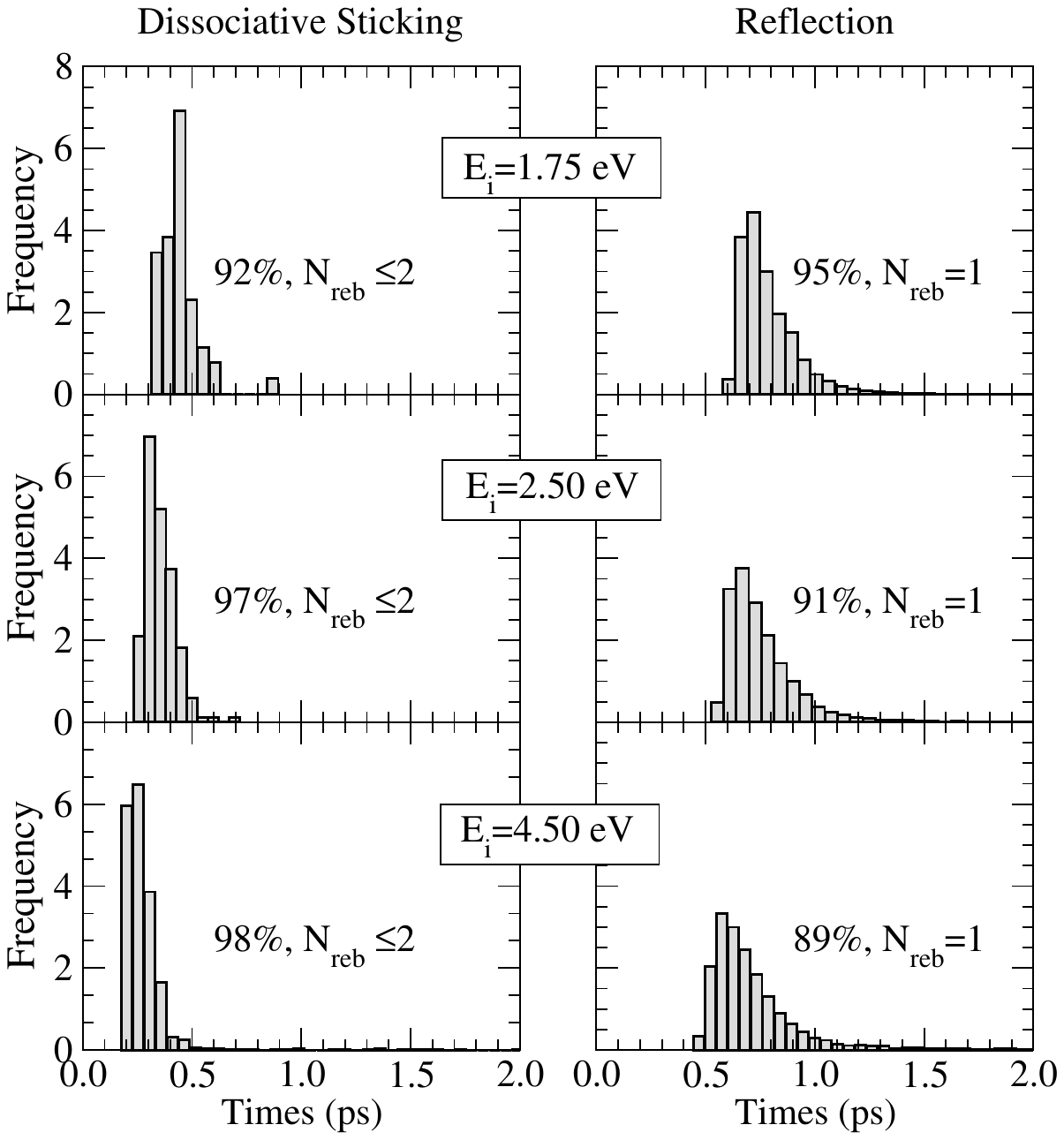}
	\caption{Times for dissociative sticking (left panels) and reflection (right panels) of CO$_{2}$ molecules in a (6$\times$6) supercell at $\mathrm{T_{s}}$=300 K. Three different values of the initial translational kinetic energy of the molecule are evaluated: $\mathrm{E_{i}}$=1.75 eV (upper panels), $\mathrm{E_{i}}$=2.50 eV (middle panels), $\mathrm{E_{i}}$= 4.50 eV (lower panels).}
	\label{fig:Times}
\end{figure}

In the $\mathrm{E_{i}}$-range in which P$_{\mathrm{diss}} >$ $\mathrm{P_{molads}}$ (i.e. $\mathrm{E_{i}} \geq$ 1.75 eV), both dissociative sticking and reflection 
events take place with interaction times smaller than 
0.6 ps and 1.2 ps respectively.
This is illustrated by Fig.\ \ref{fig:Times}.
In addition, it shows that for 92-98\% (89-95\%) of reactive (reflected) trajectories, dissociation (reflection) takes place after less than three rebounds, i.e.\ N$_{\mathrm{reb}} \leq 2$ (after only one rebound, i.e.\ N$_{\mathrm{reb}}$=1).
Similar short interaction times and small number of rebounds have also been
found for $\mathrm{E_{i}}$=1.25 eV and 1.50 eV but in these cases,
more trajectories would be needed to perform an analysis of
the distribution of dissociation times with small statistical uncertainties
and so, these energies have not been considered in Fig.\ \ref{fig:Times}.
These results clearly show that for $\mathrm{E_{i}} \gtrsim$ 1.25 eV for which
we have found dissociative sticking events, the scattering of most molecules
is direct no matter whether they reflect or dissociate on Cu(110).

\begin{figure}
	\includegraphics[width=0.4\linewidth]{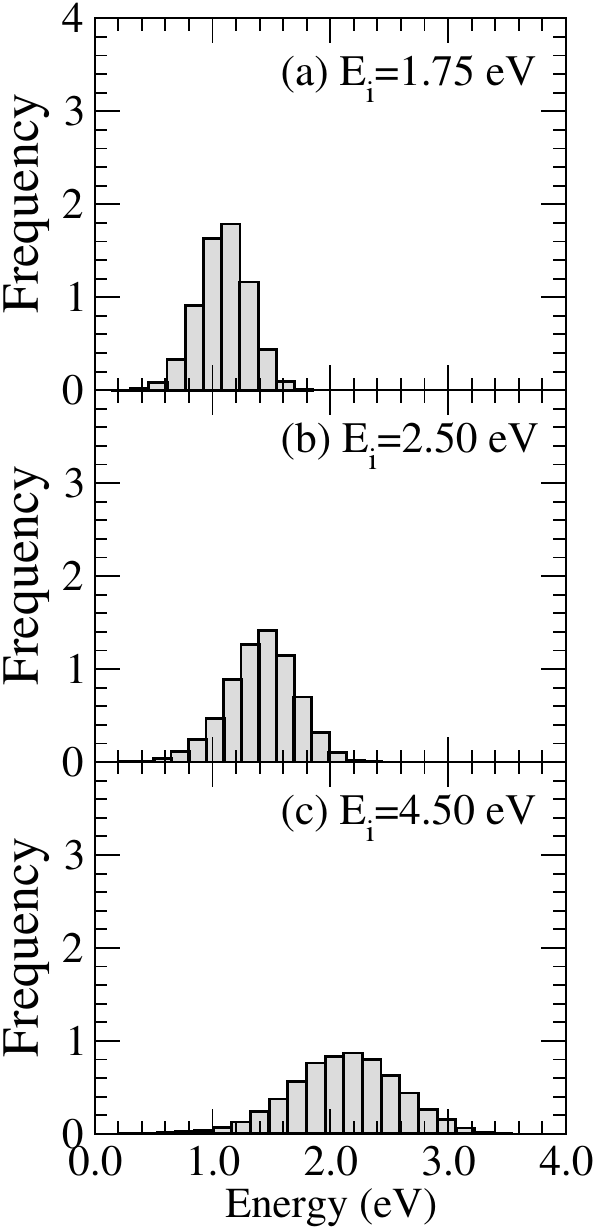}
	\caption{Distribution of energy transferred by reflected CO$_{2}$ molecules to the Cu(110) surface at $\mathrm{T_{s}}$ =300 K for normal incidence. (a) $\mathrm{E_{i}}$=1.75 eV, (b) $\mathrm{E_{i}}$ =2.50 eV, (c) $\mathrm{E_{i}}$= 4.50 eV.}
	\label{fig:reflection-energy-transfer}
\end{figure}

In spite of the single-rebound character of reflection events for most molecules, they transfer a significant fraction of the initial translational 
energy to the surface. Fig.\ \ref{fig:reflection-energy-transfer}
shows that for $\mathrm{E_{i}}$=1.75 eV, 2.5 eV and 4.5 eV, 
reflected molecules transfer to the surface between 25\% and 85\% of  their initial translational energy.
Interestingly, even the upper bound of the energy transfer
is smaller than the value predicted by the 
Baule formula \cite{Harris1991a}, $\Delta \mathrm{E}=4\,\alpha\,{\mathrm{E_{i}}}/(1+\alpha)^2 \simeq$ 0.97 $\mathrm{E_{i}}$, with $\alpha \simeq 0.69$ being the ratio of the mass of a CO$_2$ molecule and the mass of a Cu atom.

\begin{figure}
%	\centering     
	\includegraphics[width=0.7\linewidth]{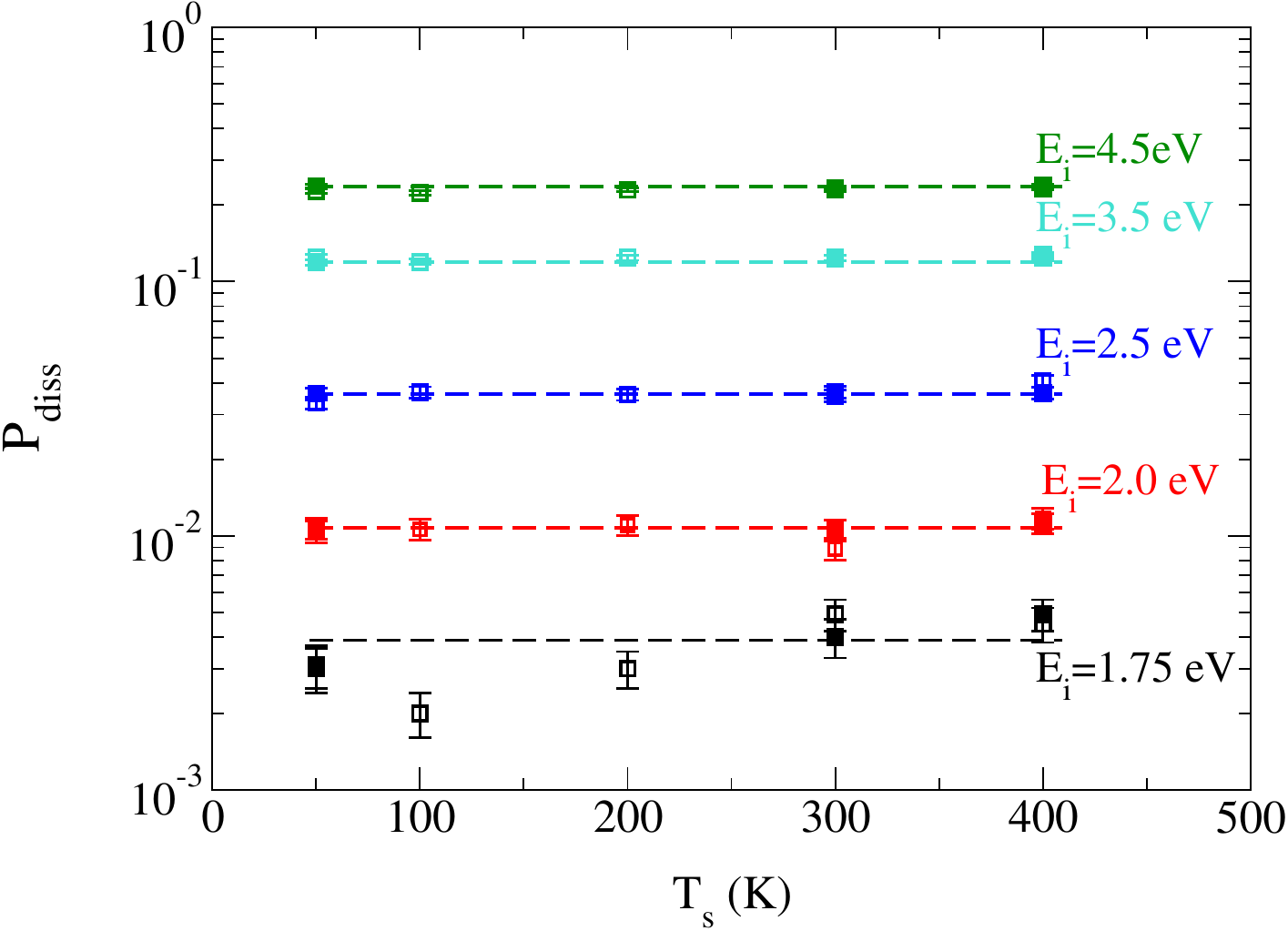}
	\caption{$\mathrm{P_{diss}}$ as a function of surface temperature, $\mathrm{T_{s}}$, for
	normal incidence and $\mathrm{E_{i}}$: 1.75 eV (black), 2.0 eV (red), 2.5 eV (blue),
	3.5 eV (turquoise), and 4.5 eV (dark green). Full symbols, (6$\times$6) supercell; open symbols, (3$\times$3) supercell. 
	The dashed horizontal lines are plotted to guide the eyes.}
	\label{fig:TSdisoc_prob_vs_ener}
\end{figure}

In Fig.\ \ref{fig:TSdisoc_prob_vs_ener} we consider the influence of $\mathrm{T_{s}}$ on 
$\mathrm{P_{diss}}$ between 50 K and 400 K. 
Our calculations predict a negligible $\mathrm{T_{s}}$-dependence 
in the whole range of $\mathrm{E_{i}}$ values considered, in line with experiments \cite{Singh2024a,Singh2024b}.
This is in contrast with activated dissociation of other molecule/surface systems (e.g.\ for CH$_4$ interacting with metal surfaces) for which $\mathrm{P_{diss}}$ strongly increases with $\mathrm{T_{s}}$, in particular 
at low impact energies (see e.g.\ Ref. \citenum{Jackson2013a}).
Interestingly, allowing Cu atoms relaxation provoke a 0.15 eV reduction of the energy barrier for dissociation of CO$_2$ on Cu(110) 
(with respect to the value obtained within a rigid surface model)
similar to what is found for various CH$_4$/metal surface systems\cite{Nave2010a}.
However, it must be noted that whereas the lattice distortion favorable for CH$_4$ activation simply consists in the shift up of its closest metal atom, in the case of CO$_2$, appropriate concerted displacements of four neighbor Cu atoms with respect to their equilibrium positions are needed for the 0.15 eV energy-barrier reduction mentioned above.
Therefore, the probability for a molecule to find
a particular lattice distortion more convenient for dissociation simply induced by thermal fluctuations when $\mathrm{T_{s}}$ increases, is expected to be much smaller for CO$_2$ on Cu(110) than for CH$_4$ on metal surfaces.
This provides a possible explanation of why surface temperature is not effective activating CO$_2$ in contrast with CH$_4$.

Before concluding this section it is worth commenting that 
for $\mathrm{E_{i}}$=1.5 eV we have also performed calculations of $\mathrm{P_{diss}}$
for the CO$_2$ molecules initially in four vibrationally excited states:
the first excited state of the asymmetric stretching (AS,$\nu$=1), and
symmetric stretching (SS,$\nu$=1) modes, and the first and second excited state of the bending (B,$\nu$=1,2) mode.
According to our ANN-PES, the frequencies of these three vibrational 
modes of CO$_2$ in vacuum are: 2295 cm$^{-1}$ ($\hbar\, \omega_{AS}$=0.285 eV), 
1284 cm$^{-1}$ ($\hbar\, \omega_{SS}$=0.159 eV), and
616 cm$^{-1}$ ($\hbar\, \omega_{B}$=0.076 eV) respectively, in good agreement with the experimental values.
For all these excited states, $\mathrm{P_{diss}}$ is larger than for the ground state
of CO$_2$, by factors 
$\mathrm{P_{diss}}$(AS,$\nu$=1)/$\mathrm{P_{diss}}$(GS)=4.1,
$\mathrm{P_{diss}}$(SS,$\nu$=1)/$\mathrm{P_{diss}}$(GS)=2.5,
$\mathrm{P_{diss}}$(B,$\nu$=2)/$\mathrm{P_{diss}}$(GS)=2.4, and  $\mathrm{P_{diss}}$(B,$\nu$=1)/$\mathrm{P_{diss}}$(GS)=1.8. 
The ordering of reactivity of these vibrationally excited states
is consistent with that found by Yin and Guo \cite{Yin2024},
and with the ordering of their internal energies.
In particular, it is interesting to note that 
for the excited states SS,$\nu$=1 and B,$\nu$=2, which have very similar total 
internal energies, the corresponding values of $\mathrm{P_{diss}}$ are also very close to each other, suggesting a small mode-specificity.

\subsubsection{{\em Final state} of dissociation products} \label{sec:res:postdissoc}

\begin{figure}
%\centering
\includegraphics[width=0.38\textwidth]{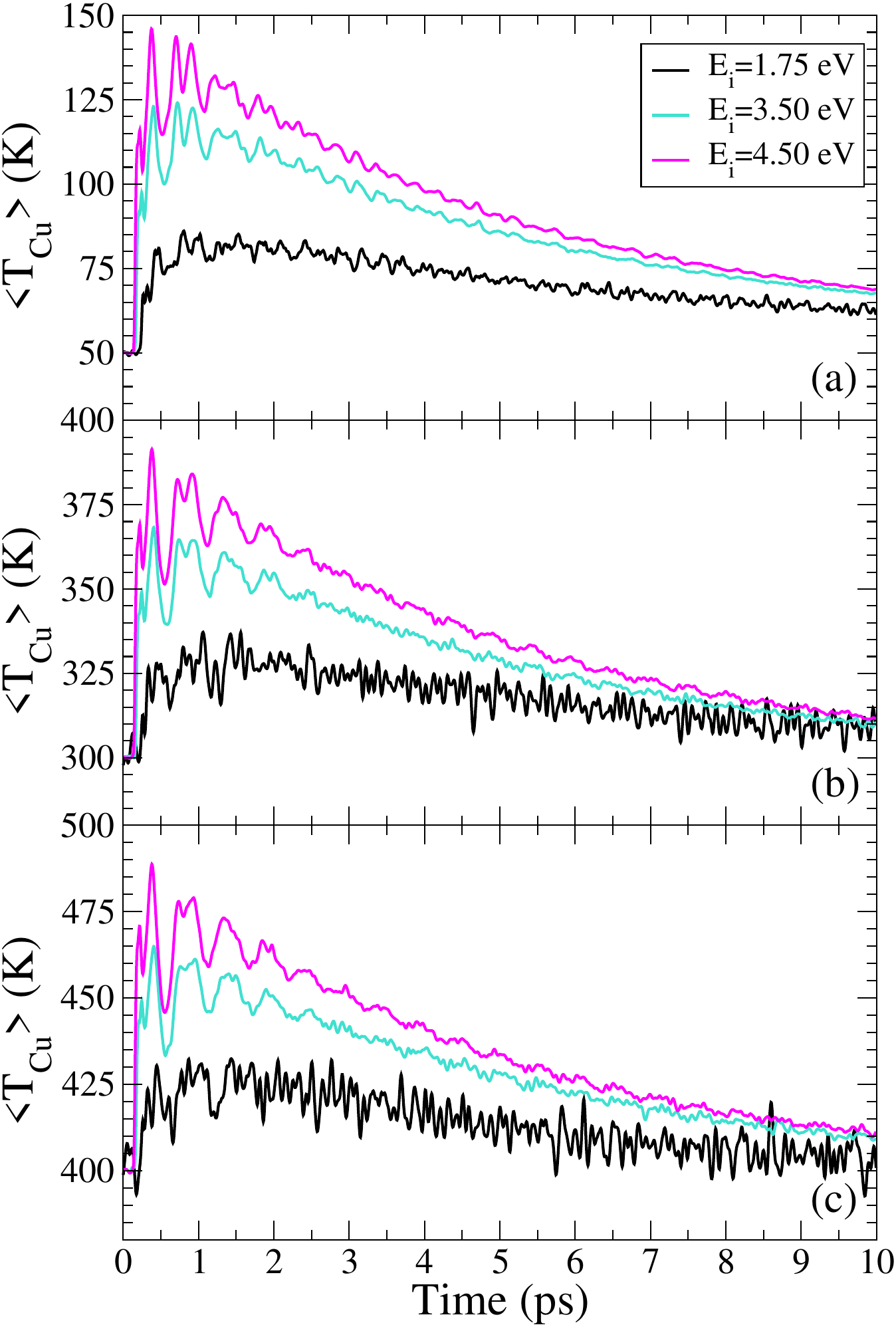}
\caption{Time evolution of the mean surface temperature $\left\langle \mathrm{T_{Cu}} \right\rangle $ for trajectories of dissociated molecules for $\mathrm{E_{i}}$=1.75 eV, 3.50 eV, and 4.5 eV, and initial surface temperatures: (a) $\mathrm{T_{s}}$=50 K, (b) $\mathrm{T_{s}}$=300 K, and (c) $\mathrm{T_{s}}$=400 K.}
\label{fig:average_temperature}
\end{figure}

To characterize the final state of the dissociation products of CO$_2$, for reactive trajectories we extended the integration time up to 10 ps. 
The choice of this integration time is justified by the temporal evolution of the average substrate temperature $<\mathrm{T_{Cu}}>$ shown in Fig.\ \ref{fig:average_temperature}.
For three impact energies of CO$_2$ ranging from $\mathrm{E_{i}}$=1.75 eV to $\mathrm{E_{i}}$=4.50 eV, and for different initial surface temperatures: $\mathrm{T_{s}}$=50 K (a), $\mathrm{T_{s}}$=300 K (b), and $\mathrm{T_{s}}$=400 K (c), $<\mathrm{T_{Cu}}>$ presents an abrupt increase at $\mathrm{t}\sim 0.2-0.3$~ps (more pronounced for higher impact energies) due to the energy transfer from the fast molecules to the surface.
After this initial increase, $<\mathrm{T_{Cu}}>$ gradually decays over time
to reach at $\mathrm{t_{max}}$=10 ps, a value similar to the initial $\mathrm{T_{s}}$.
Thus, an integration time of 10 ps ensures that most of the energy initially transferred to the substrate by the impinging molecule, has been dissipated.

\begin{figure}
	\centering
	\includegraphics[width=0.8\linewidth]{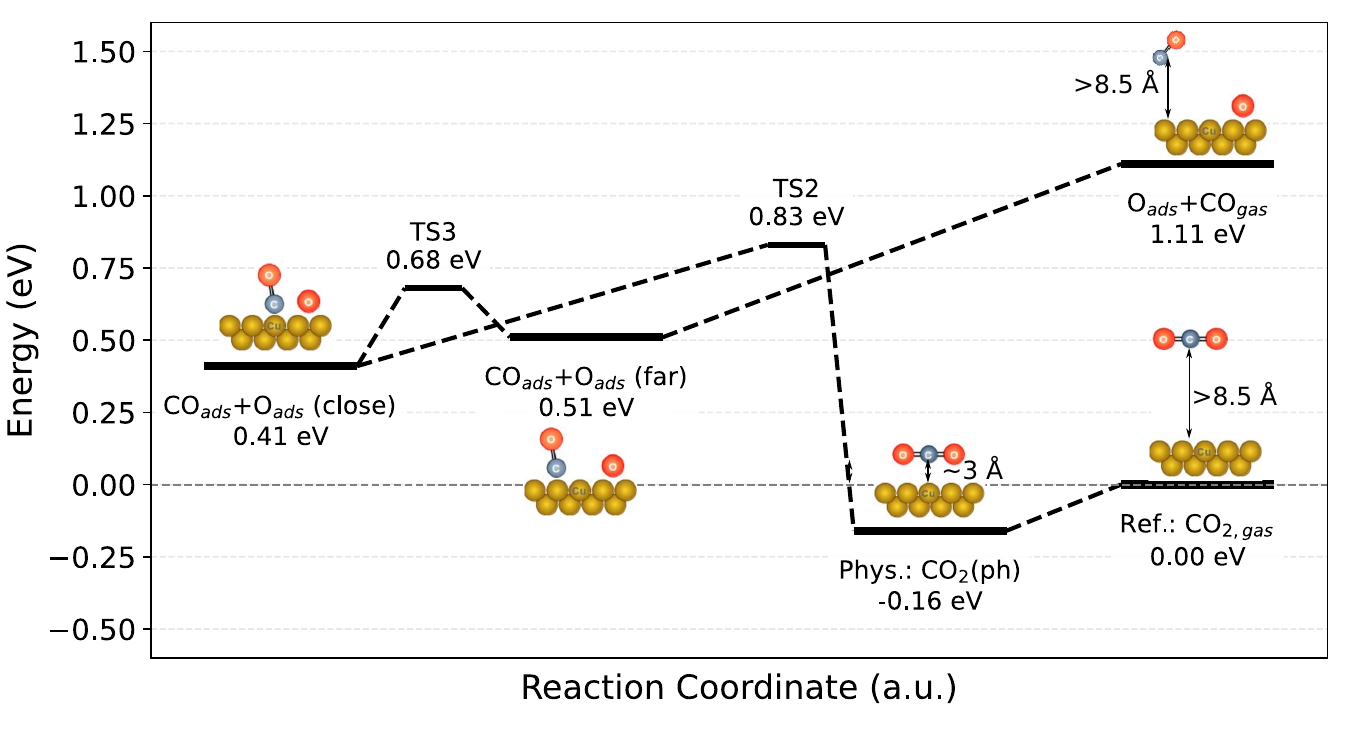}
	\caption{Minimum energy pathways of various possible processes 
		involving $\mathrm{CO_{ads}}$ and $\mathrm{O_{ads}}$ after CO$_2$ dissociation on Cu(110): diffusion of $\mathrm{CO_{ads}}$ far from 
		$\mathrm{O_{ads}}$, recombination to form CO$_2$ physisorbed or desorbed, and CO desorption.}
	\label{fig:energy_paths_post}
\end{figure}

Table \ref{FinalChannels-1} shows the percentage of the different final states of the CO$_2$ dissociation fragments on Cu(110), observed at $\mathrm{t_{max}}$= 10 ps for various initial molecular energies and surface temperatures.
For the great majority of the trajectories,
either CO and O both remain adsorbed on the surface (hereafter referred to as $\mathrm{CO_{ads}}$+$\mathrm{O_{ads}}$ states) or only the O atom remains adsorbed whereas the CO molecule desorbs (hereafter referred to as $\mathrm{CO_{gas}}$+$\mathrm{O_{ads}}$ states).
The energetics of the most probable processes that can take place
after dissociation starting from the lowest energy configuration with $\mathrm{CO_{ads}}$ and $\mathrm{O_{ads}}$ close to each other (characterized by a potential energy 0.41 eV) are summarized in Fig.\  \ref{fig:energy_paths_post}.
Recombination events leading to desorbed or physisorbed CO$_2$ molecules (i.e.\ $\mathrm{CO_{2,gas}}$ or $\mathrm{CO_{2}(ph)}$ respectively) are very scarce. 
These final states have been found only for $\sim 1$\% of the dissociated molecules in the most favorable condition of high $\mathrm{E_{i}}$ and $\mathrm{T_{s}}$ values. 
Due to this reason, the percentage of recombination events for each initial condition ($\mathrm{E_{i}}$, $\mathrm{T_{s}}$) have not been included in Tables \ref{FinalChannels-1} and \ref{FinalChannels-2} (see below).

\begin{table}[htb!]
	\begin{center}
		\caption{Percentage of different final states post-dissociation of CO$_2$ on Cu(110) at the final integration time $ \mathrm{t_{max}}$=10 ps, for $\mathrm{T_{s}}$= 50, 300, and 400 K and $\mathrm{E_{i}}$=1.75, 2.00, 2.50, 3.50, and 4.50 eV. The results correspond to calculations performed using a 6x6 supercell. The definition of each type of final states, is provided in the main text.} 
		\label{FinalChannels-1}
		%\begin{tabular}{| c | S[table-format=2.2] | S[table-format=2.2]  | S[table-format=2.2] | c | S[table-format=2.2] | S[table-format=2.2] |}
		\begin{tabular}{ | c | c | c | c |}
			\hline
			\textbf{$\mathrm{\bf E_{i}}$ (eV)} & \textbf{$\mathrm{\bf CO_{gas}}$+$\mathrm{\bf O_{ads}}$} & \textbf{$\mathrm{\bf CO_{ads}}$+$\mathrm{\bf O_{ads}}$} & \textbf{+Cu-adatom} \\ 
			\hline
			\hline
			\multicolumn{4}{|c|}{$\mathrm{T_{s}}$=50 K} \\ \hline 
			\textbf{1.75} &  0.00 & 100.00 (90.00) & 0.00  \\ \hline
			\textbf{2.00} &  0.96 & 95.19 (86.54) & 3.85  \\ \hline 
			\textbf{2.50} &  8.84 & 87.56 (68.78) & 3.59  \\ \hline 
			\textbf{3.50} &  31.20 & 60.89 (38.10) & 7.23   \\ \hline 
			\textbf{4.50} &  55.84 & 30.88 (14.96)  & 12.36  \\ 
			\hline
			\hline
			\multicolumn{4}{|c|}{$\mathrm{T_{s}}$=300 K} \\ \hline
			\textbf{1.75} &  0.00 & 97.50 (85.00)  & 2.50  \\ \hline
			\textbf{2.00} &  2.86 & 89.53 (75.24) & 7.62  \\ \hline 
			\textbf{2.50} &  14.17 & 74.94 (51.23) & 10.08  \\ \hline 
			\textbf{3.50} &  42.88 & 44.74 (22.25)  & 11.33  \\ \hline 
			\textbf{4.50} &  56.69 & 27.60 (10.42)  & 14.51  \\ 
			\hline
			\hline			 
			\multicolumn{4}{|c|}{$\mathrm{T_{s}}$=400 K} \\ \hline
			\textbf{1.75} &  2.04 & 81.63 (53.06) & 16.33  \\ \hline
			\textbf{2.00} &  3.54 & 76.39 (59.29) & 19.46  \\ \hline 
			\textbf{2.50} &  16.30 & 66.03 (39.67) & 16.58  \\ \hline 
			\textbf{3.50} &  40.53 & 39.09 (18.23) & 19.02   \\ \hline 
			\textbf{4.50} &  56.15 & 21.79 (14.69) & 19.72   \\ \hline
		\end{tabular}
	\end{center}
\end{table}

The results reported in Table \ref{FinalChannels-1} show that $\mathrm{CO_{ads}}$+$\mathrm{O_{ads}}$ is the dominant type of final dissociated states at the lowest impact energies considered, regardless of the value of $\mathrm{T_{s}}$. 
When $\mathrm{E_{i}}$ and/or $\mathrm{T_{s}}$ increase, the percentage of $\mathrm{CO_{ads}}$+$\mathrm{O_{ads}}$ ($\mathrm{CO_{gas}}$+$\mathrm{O_{ads}}$) final states decreases (increases) significantly. 
In the $\mathrm{CO_{ads}}$+$\mathrm{O_{ads}}$ column of Table \ref{FinalChannels-1}, the first number indicates the total percentage of this type of final state whereas the second value (between parentheses) refers only to the cases in which both fragments remain  within the same surface unit cell 
($\mathrm{|X_C-X_{O_{ads}}|} \leq$ 2.654 \AA\ and $\mathrm{|Y_C-Y_{O_{ads}}|} \leq $ 3.754 \AA).
At low impact energies, the majority of the $\mathrm{CO_{ads}}$ and $\mathrm{O_{ads}}$ fragments remain close to each other within the same Cu(110) unit cell, but they end up farther apart from each other when $\mathrm{E_{i}}$ increases.
For instance, for $\mathrm{T_{s}}$=300 K, 85\% of dissociation products remain adsorbed close to each other for $\mathrm{E_{i}}$=1.75 eV but only $\sim 10$\% do it for $\mathrm{E_{i}}$=4.5 eV. 
This indicates that the initial translational energy of CO$_2$ does not dissipate efficiently before dissociation, being then transferred to a large extent, to the relative motion of the dissociation products. 

The dissociation products of type $\mathrm{CO_{ads}}$+$\mathrm{O_{ads}}$ that end far from each other after 10 ps are not likely to recombine and for $\mathrm{T_{s}}$ above $\sim 200$~K (the desorption temperature of CO \cite{Horn1977,Christiansen1992,Ahner1996}), it is expected that most of them will eventually convert into $\mathrm{CO_{gas}}$+$\mathrm{O_{ads}}$ due to CO thermal desorption. 
In contrast, one might argue that those $\mathrm{CO_{ads}}$+$\mathrm{O_{ads}}$ dissociation products that after 10 ps are still close to each other might eventually recombine and desorb as CO$_2$. This could approach the theoretical
$\mathrm{P_{diss}}$ to experiments at high energies (see Fig.\ \ref{fig:stick_GS_prob}).
However, this is not the case since for 
$\mathrm{E_{i}}$=4.5 eV and $\mathrm{T_{s}}$=300 K, only $\sim 10$\% of the 
$\mathrm{CO_{ads}}$+$\mathrm{O_{ads}}$ dissociation products remain close to each other and even if all of them recombine, this would entail only a 10\% reduction of $\mathrm{P_{diss}}$.

It is important to mention that in the columns of Table \ref{FinalChannels-1} labeled as $\mathrm{CO_{gas}}$+$\mathrm{O_{ads}}$ and $\mathrm{CO_{ads}}$+$\mathrm{O_{ads}}$, we have only considered
final states of the dissociation products in which the surface structure is
preserved. 
In contrast, in the right column of Table \ref{FinalChannels-1} labeled as {\em +Cu-adatom}, we report the total percentage of final states of the dissociation products that have been found coexisting with one Cu adatom located above the topmost-layer of the Cu(110) surface. 
The presence of Cu adatoms in the final state of each trajectory was 
evaluated considering the final height of each Cu atom with respect to the 
average height of the Cu atoms in the topmost-layer of Cu(110) in the initial configuration of the trajectory. 
We consider that a Cu atom is such an adatom when, at $\mathrm{t_{max}}$=10 ps, its $\mathrm{Z}$ coordinate, $\mathrm{Z_{Cu}}>\mathrm{\bar{Z}_{Cu-top}^{ini}}+0.5\,\mathrm{d_{110}}$, being $\mathrm{\bar{Z}_{Cu-top}^{ini}}$ the average $\mathrm{Z}$ coordinate of all the topmost-layer Cu atoms in the initial state of the trajectory, and $\mathrm{d_{110}}$ the ideal (bulk-truncated) inter-layer distance between consecutive (110) crystallographic planes.
Other types of lattice distortions have also been found in the final states of dissociation products but much less frequently than those with Cu adatoms, and so we will disregard them in what follows.

In Table \ref{FinalChannels-1} it is shown that the percentage of {\em +Cu-adatom} final states of dissociation products is negligible at low impact energies, increases with $\mathrm{E_{i}}$ and $\mathrm{T_{s}}$, and reaches its maximum value $\sim 20$\% for $\mathrm{E_{i}}$=4.5 eV and $\mathrm{T_{s}}$=400 K.
For final states of type {\em +Cu-adatom} we can also separate those configurations where both CO and O are adsorbed on the surface on one side, and those in which only the O atom remains adsorbed on the surface due to the desorption of CO on the other.
These final states might also be referred to as $\mathrm{CO_{ads}}$+$\mathrm{O_{ads}}$ and $\mathrm{CO_{gas}}$+$\mathrm{O_{ads}}$ respectively as it was done before.
However, to emphasize that now the structures involve one Cu adatom
we introduce a new specific notation:
\begin{itemize}
	\item {$\mathrm{\bf CO_{gas}}$+($\mathrm{\bf O_{ads}}$+{\bf Cu})}: states in which CO is desorbed and 
	the O atom remains adsorbed on the surface near (i.e.\ to a distance shorter than the length of the diagonal of the unit cell of Cu(110), 4.6 \AA) or far from the Cu adatom,
	\item {$\mathrm{\bf CO_{ads}}$+{\bf Cu}+$\mathrm{\bf O_{ads}}$}: states in which both CO and O are adsorbed on the surface being at most one of them close to the Cu adatom, 
	\item {\textbf{(O-Cu-CO)}$\mathrm{\bf_{ads}}$}: states in which both CO and O are adsorbed on the surface close to the Cu adatom, forming an almost linear O-Cu-CO chain moiety anchored to the surface through the O atom and being the Cu atom slightly detached from the surface.
\end{itemize}

\begin{table}[htb!]
	\begin{center}
		\caption{Classification of final-state channels involving Cu atoms that have shifted up with respect to the topmost layer of the slab, resulting from CO$_{2}$ dissociation on a 6$\times$6-Cu(110) cell, for $\mathrm{T_{s}}$=50 K, 300 K, and 400 K. A detailed description of each configuration is provided in the main text.}
		\label{FinalChannels-2}
		\begin{tabular}{ | c | c | c | c |}
			\hline
			$\mathrm{\bf E_{i}}$ \textbf{(eV)} &  $\mathrm{\bf CO_{gas}}$+($\mathrm{\bf O_{ads}}+{\bf Cu}$)  &
			$\mathrm{\bf CO_{ads}}+{\bf Cu}+\mathrm{\bf O_{ads}}$ &
			\textbf{(O-Cu-CO)}$\mathrm{\bf _{ads}}$   \\
			\hline
			\hline
			\multicolumn{4}{|c|}{$\mathrm{T_{s}}$=50 K} \\ \hline
			\textbf{1.75} &  0.00 (0.00) &  0.00  (0.00, 0.00 )  & 0.00     \\ \hline
			\textbf{2.00} &  0.00 (0.00) & 0.00  (0.00, 0.00) & 3.85  \\ \hline
			\textbf{2.50} &  0.28 (0.28) & 0.28  (0.00, 0.28) & 3.04   \\ \hline
			\textbf{3.50} &  1.60 (1.60) & 2.86  (0.93, 1.60) & 2.78  \\ \hline
			\textbf{4.50} &  6.05 (5.97) & 4.00 (1.01,  2.82) & 2.27  \\ \hline
			\hline
			\multicolumn{4}{|c|}{$\mathrm{T_{s}}$=300 K} \\ \hline
			\textbf{1.75} &  0.00 (0.00) & 0.00  (0.00, 0.00) & 2.50 \\ \hline
			\textbf{2.00} &  0.00 (0.00) & 1.90  (1.90, 0.00) & 5.71 \\ \hline
			\textbf{2.50} &  0.27 (0.27) & 2.99 (1.09, 0.82 ) & 6.81
			\\ \hline
			\textbf{3.50} &  2.51 (1.94) & 4.12 (1.78, 1.29) & 4.61 \\ \hline
			\textbf{4.50} &  5.68 (4.34) & 5.83 (1.77, 2.68) & 2.92 \\ \hline
			\hline
			\multicolumn{4}{|c|}{$\mathrm{T_{s}}$=400 K} \\ \hline
			\textbf{1.75} &  0.00 (0.00) & 10.20 (10.20, 0.00 ) & 6.12 \\ \hline
			\textbf{2.00} &  0.00 (0.00) & 11.49 (7.96, 0.88 ) & 7.96 \\ \hline
			\textbf{2.50} &  0.82 (0.27) & 9.51 (4.62, 3.26 ) & 6.25 \\ \hline
			\textbf{3.50} &  3.43 (2.48) & 8.88 (3.52, 2.64 ) & 6.63 \\ \hline
			\textbf{4.50} &  7.67 (5.03) & 7.84 (2.60, 2.64 ) & 4.17 \\ \hline
		\end{tabular}
	\end{center}
\end{table}

\begin{figure}
	\centering
	\includegraphics[width=1\linewidth]{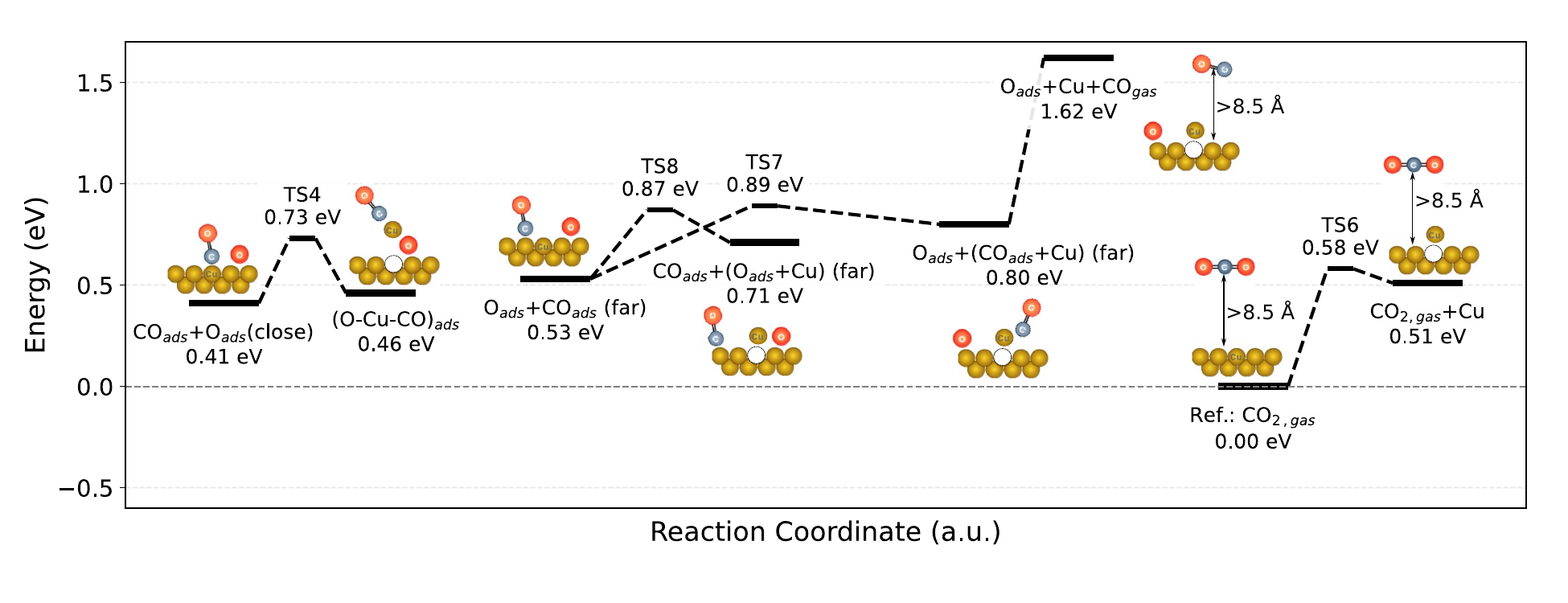}
	\caption{Minimum energy pathways of various possible processes 
		involving $\mathrm{CO_{ads}}$ and $\mathrm{O_{ads}}$ after CO$_2$ dissociation on Cu(110), in presence of and/or inducing Cu adatom-vacancy pair formation.}
	\label{fig:energy_paths_post-Cu-adatoms}
\end{figure}

In Table \ref{FinalChannels-2} we report the percentages of these three types of final states of dissociation products found in our calculations.
In the case of $\mathrm{CO_{gas}}$+($\mathrm{O_{ads}}$+Cu) we report the total
percentage of this type of states, indicating between parentheses the
percentage in which the O atom is close to the Cu adatom (i.e.\ to a distance shorter than 4.6 \AA). 
It is observed that the percentage of this type of final state
increases with the initial impact energy of the molecules.
In most cases $\mathrm{O_{ads}}$ is close to the Cu adatom suggesting that the O atom
stabilizes or even assists the formation of the Cu adatom.
The energetics of the creation of a single Cu adatom (predicted by our ANN-PES) for different relative distances between $\mathrm{O_{ads}}$ and $\mathrm{CO_{ads}}$ from the Cu adatom is shown in Fig.\ \ref{fig:energy_paths_post-Cu-adatoms}.
It can be observed that the energy cost and the barrier for the formation of a single Cu adatom with no influence of $\mathrm{CO_{ads}}$ and $\mathrm{O_{ads}}$ are 0.51 eV and  0.58 eV (TS6) respectively.
In contrast, if $\mathrm{O_{ads}}$ is close to the Cu adatom these values reduce to 
0.71 eV - 0.53 eV = 0.18 eV and 0.87 eV - 0.53 eV = 0.34 eV (TS8) 
respectively. This is consistent with the fact that in most of the final structures $\mathrm{CO_{gas}}$+($\mathrm{O_{ads}}$+Cu), the $\mathrm{O_{ads}}$ species is close to the Cu adatom.

In the column labeled as $\mathrm{CO_{ads}}$+Cu+$\mathrm{O_{ads}}$ 
we report the total percentage of this type of final dissociated states
followed by two numbers between parentheses,
corresponding to (from left to right) 
configurations with
$\mathrm{CO_{ads}}$ close to (and $\mathrm{O_{ads}}$ far from) the Cu adatom, and 
$\mathrm{O_{ads}}$ close to (and $\mathrm{CO_{ads}}$ far from) the Cu adatom.
Here, we have considered that a dissociation product ($\mathrm{CO_{ads}}$ or $\mathrm{O_{ads}}$) is close to (far from) a Cu adatom when it is located a distance  $\leq$ 4.6 \AA\ ($\geq$ 4.6 \AA) to it.
First, as expected, total percentage of this type of final states involving a Cu adatom increases with surface temperature.
In addition, in most of $\mathrm{CO_{ads}}$+Cu+$\mathrm{O_{ads}}$ final states,
either $\mathrm{CO_{ads}}$ or $\mathrm{O_{ads}}$ are close to the Cu adatom which indicates 
that the presence of $\mathrm{CO_{ads}}$ also stabilizes and/or assists the formation of Cu adatoms. This is confirmed by  Fig.\ \ref{fig:energy_paths_post-Cu-adatoms} where we show that
if $\mathrm{CO_{ads}}$ is close to the Cu adatom the energy cost and barrier 
for formation of a Cu adatom are
0.80 eV - 0.53 eV = 0.27 eV and 0.89 eV - 0.53 eV = 0.36 eV (TS7) 
(to be compared with 0.51 eV and 0.58 eV for the {\em non-adsorbate-assisted} creation of a Cu adatom) respectively.
Moreover, it is important to mention that in most of the final states 
with $\mathrm{CO_{ads}}$ close to the Cu adatom, the molecule is directly chemisorbed
on top of it, the desorption energy from this state being 0.82 eV, i.e.\
0.22 eV higher than the desorption energy of CO from Cu(110) with no adatoms.

Finally, the percentage of (O-Cu-CO)$\mathrm{_{ads}}$ final states also 
increases with surface temperature.
The energy of this final state is only 0.05 eV higher than the most stable 
final dissociated state with $\mathrm{CO_{ads}}$ and $\mathrm{O_{ads}}$ chemisorbed
within  the same unit cell of Cu(110) and the energy barrier for the creation of this linear  O-Cu-CO chain moiety (TS4) is only 0.73 eV - 0.41 eV = 
0.32 eV (Fig.\ \ref{fig:energy_paths_post-Cu-adatoms}).
This explains the relatively large probabilities of this unexpected final state in which a Cu adatom looks detached from the surface and forming a 
O-Cu-CO chain anchored to the surface through the dissociated O atom.
It might be argued that this final state might be simply a consequence of an artifact of the ANN-PES used in the QCT calculations. 
However, it is not the case.
It must be emphasized that some of this kind of (O-Cu-CO)$\mathrm{_{ads}}$ structures are included in the data base used
for training of the ANN-PES during the active learning procedure employed. 
The MAE of the ANN-PES for a sub-set of 25 of these configurations is only 1.9 meV/atom.
This highlights the importance of using the active learning method 
that allows unexpected configurations visited by the trajectories during the simulations (that would be otherwise overlooked), to be identified and incorporated into the training database.

\section{Conclusions}

In this work we have used quasi-classical trajectory calculations
to investigate the dynamics of CO$_2$ molecular and dissociative adsorption on Cu(110) as a function of the initial translational energy of the molecule, $\mathrm{E_{i}}$, and surface temperature, $\mathrm{T_{s}}$.
These calculations were performed by using an artificial neural network (ANN) potential energy surface (PES) optimized in this work for the CO$_2$/Cu(110) system, through an iterative active learning approach, to properly describe density functional theory (DFT) total energies for a large set of system configurations.
By using the vdW-DF2 exchange-correlation functional, we obtained
that the global energy minimum of the system corresponds
to weakly bound CO$_2$ physisorbed on Cu(110) with energy
$\sim 0.17$~eV below the lowest value obtained for the molecule far from the surface that we take as our zero-energy reference level.
We have also found chemisorbed bent state with energy
+0.49 eV, which is highly unstable because an energy barrier of
only +0.01 eV is found along the minimum energy path (MEP) connecting this local minimum with the global minimum of the PES.
In addition, the energy of the transition state found along the
MEP connecting the physisorbed state and the most stable dissociated state (of energy +0.41 eV) corresponding to both CO and O adsorbed within the same surface unit cell is +0.83 eV.

We have found that physisorption is the dominant adsorption channel below the desorption temperature of CO$_2$ (i.e. $\mathrm{T_{s}} \sim 90$~K) and for $\mathrm{E_{i}}$ up to $\sim 1.75$~eV also in qualitative agreement with experiments but the initial decrease of the corresponding  $\mathrm{P_{molads}}$($\mathrm{E_{i}}$) theoretical curve is much
more pronounced than in experiments.
The dissociation probability presents a monotonic increasing $\mathrm{E_{i}}$-dependence with a decreasing slope of the $\mathrm{P_{diss}}$($\mathrm{E_{i}}$) curve in agreement with experiments. 
However, the agreement with supersonic molecular beam experiments is only qualitative since our results underestimate the experimental values for
$\mathrm{E_{i}} \lesssim$ 2 eV and 
overestimate them for $\mathrm{E_{i}} \gtrsim$ 2 eV.

Also in agreement with experiments, we have found that $\mathrm{P_{diss}}$
is not affected by surface temperature between 50 K and 400 K.
Interestingly, this is in spite of a non-negligible reduction
of the energy barrier for CO$_2$ when Cu atoms are allowed to adapt their optimum positions in presence of the molecule. 
This negligible $\mathrm{T_{s}}$-sensibility is attributed to the fact that
concerted particularly appropriate  displacements of various
neighboring Cu atoms is required to significantly reduce the energy of the CO$_2$ transition state for dissociation. 
Therefore, the probability of inducing such a particular set of
Cu-displacements thanks to thermal fluctuations by increasing surface temperature, in order to offer to the incoming (fast) molecules more favorable dissociation pathways is very small.

We have also characterized the final state of the dissociation 
products, the most probable one being $\mathrm{CO_{ads}}$+$\mathrm{O_{ads}}$
(both species adsorbed) at low energies and $\mathrm{CO_{gas}}$+$\mathrm{O_{ads}}$
(only O adsorbed and CO desorbed) at higher energies.
However, above 200 K (the desorption temperature of CO from Cu(110))
it is expected that the $\mathrm{CO_{ads}}$ species found in simulations
would eventually desorb at longer times, 
remaining at the end only oxygen atoms adsorbed on the surface in agreement with experiments performed for instance at room temperature (RT).
Interestingly, we have found that for molecules with impact energies above $\sim 2.5$~eV (in particular for RT and above),
dissociation events produce with a non-negligible probability, 
surface distortions involving Cu adatoms. 
The presence of the CO and O adsorbed species reduce the energy
barrier for creation of Cu adatoms which in turn can be overcome
more easily by the system for high impact energy molecules due to 
a relatively efficient energy transfer to lattice vibrations.
The changes observed at $\mathrm{E_{i}} \sim 2.5$~eV, in the final state of dissociation products from the case of low to high energy molecules
might provide a possible explanation for the experiments that 
have found that the saturation atomic oxygen coverage 
achieved for CO$_2$ with impact energies larger than 3 eV (0.66 ML)
is larger than for lower energy molecules (0.50 ML).
Still, additional experimental and also theoretical investigations
are needed to fully elucidate the origin of the latter so far unexplained experimental results.

%\section*{Supporting Information}

%There is no Supporting Information.

\begin{acknowledgments}
This work has been supported by 
the ANPCyT Project PICT-2021-I-A-01135, 
CONICET Project PIP 1679, and the UNR Project PID 80020190100011UR (Argentina). The authors acknowledge computer time provided by 
CCT-Rosario Computational Center, member of the High Performance Computing National System of Argentina (SNCAD).
\end{acknowledgments}

\section*{Author Contributions}
All the authors participated in the calculations, analysis and discussion of the results, 
and writing of the manuscript.

\section*{Data Availability Statement}
The data that support the findings of
this study are available from the
corresponding author upon reasonable
request.

\section*{Conflicts of Interest}
The authors have no conflicts to disclose.

\bibliographystyle{unsrt}
\bibliography{References}

\end{document}